\documentclass[11pt]{article} 

\usepackage[top=2cm,bottom=2cm,left=2.5cm,right=2.5cm]{geometry}
\usepackage{lscape}
\usepackage{titlesec}
\titleformat{\section}{\large\bfseries}{\thesection}{1em}{}
\titleformat{\subsection}[block]{\itshape}{\normalfont\bfseries Observation \thesubsection}{1em}{}
\usepackage{enumitem}

\usepackage{multirow}
\usepackage{multicol}

\usepackage[nodisplayskipstretch]{setspace} 
\spacing{1.25}
\usepackage{graphicx, psfrag}
\usepackage{caption}
\usepackage{subfigure}
\usepackage{booktabs}
\usepackage{fancyhdr}
\usepackage[table]{xcolor}
\usepackage[breaklinks]{hyperref}
\hypersetup{colorlinks,citecolor=blue} 
\definecolor{myred}{RGB}{233,51,11}

\definecolor{color3}{RGB}{37, 150, 190}

\usepackage{amsmath}
\usepackage{amssymb}
\usepackage{appendix}
\usepackage[round]{natbib}
\makeatletter 
\let\@fnsymbol\@arabic
\makeatother
\newcommand{\key}[2]{{\small\noindent
\textbf{#1}: #2}}
\usepackage{framed}
\usepackage{cancel}
\newcommand\blfootnote[1]{%
	\begingroup
	\renewcommand\thefootnote{}\footnote{#1}%
	\addtocounter{footnote}{-1}%
	\endgroup
}

\title{
	\Large Risk-Adjusted Valuation for Real Option Decisions
}
\author{
	\normalsize Carol Alexander$^*$\quad Xi Chen$^\dagger$\quad Charles Ward$^\ddagger$}
\date{~}

\begin{document}
\maketitle\vspace{-48pt}
\noindent\rule{\textwidth}{0.4pt}
\begin{abstract}
	\noindent We model investor heterogeneity using different required returns on an investment and evaluate the impact on the valuation of an investment. By assuming no disagreement on the cash flows, we emphasize how risk preferences in particular, but also the costs of capital, influence a subjective evaluation of the decision to invest now or retain the option to invest in future.  We propose a risk-adjusted valuation model to facilitate investors' subjective decision making, in response to the market valuation of an investment opportunity. The investor's subjective assessment arises from their perceived misvaluation of the investment by the market, so  projected cash flows are discounted using two different rates representing the investor's and the market's view. This liberates our model from perfect or imperfect hedging assumptions and instead, we are able to illustrate the hedging effect on the real option value when perceptions of risk premia diverge. During crises periods,  delaying an investment becomes more valuable as the idiosyncratic risk of future cash flows increases, but the decision-maker may rush to invest too quickly  when the risk level is exceptionally high.  Our model  verifies features established by classical { real-option valuation models} and provides many new insights about the importance of modelling divergences in decision-makers risk premia, especially during crisis periods. { It also has many practical advantages because it requires no more parameter inputs than basic discounted cash flow approaches, such as the marketed asset disclaimer method, but the outputs are much richer. They allow for complex interactions between cost and revenue uncertainties as well as an easy exploration of the effects of hedgeable and un-hedgeable risks on the real option value. Furthermore, we provide fully-adjustable Python code in which all parameter values can be chosen by the user.}
	
\end{abstract}

\key{Keywords}{Risk-adjusted discount rate, investor heterogeneity, idiosyncratic risk, internal rate of return, hedging}\\[6pt]
\key{JEL Classification}{G11, G13, G32}	\\[3pt]
\rule{\textwidth}{0.4pt}

\blfootnote{~Declarations of interest: None.}
\blfootnote{$^*$Professor of Finance, University of Sussex Business School, UK. Email: c.alexander@sussex.ac.uk. Tel: +44-1273-873950  (Corresponding Author).}
\blfootnote{$^\dagger$Lecturer in Finance, University of Sussex Business School, UK. Email: xi.chen@sussex.ac.uk. Tel: +44-1273-872703.}
\blfootnote{$^\ddagger$Professor of Property Investment and Finance, University of Reading, UK. Email: c.ward@icmacentre.ac.uk. Tel: +44-1183-788239.}
\thispagestyle{empty}
\newpage
\setcounter{page}{1}
\section{Introduction}
Investors are drastically different in their risk preferences, beliefs, goals and experiences in financial markets.
Looking back on the 2008--2009 financial crisis, a large body of research now emphasises the impact of investor heterogeneity as  an important driver of financial returns  \citep{huang2010financial,chiarella2015fear}.  Heterogeneity even leads the fluctuations in aggregate market investments \citep{baker2016disagreement}, especially during the 2008 crisis. Thus, when analysing investment opportunities, it is not enough to consider their market valuation alone. It is also vital to model the investors' subjective decision making, both theoretically and practically. 

Following \citet{M1977} a wide a variety of models can now value investment decision opportunities that are contingent on non-tradable assets or projects. Such decisions are termed  `real options'
 and there is a very large body of research that proposes different methods to value them. Real option valuation (ROV) models differ in both assumptions and ease of implementation. Classic models assume the option value is the same for all investors, and its replication in a complete market implies that the value is subsumed into the project's market price. That is, assuming there is an efficient market for a company, its market value will incorporate the value of its real options. This assumption is made by  \citet{M1977,BGN1999,B2006,SN2018,AP2019}
and many others. A further assumption of market completeness yields the company's value as the sum of the book value (the value of its assets-in-place) and a ROV (that managers create by exploiting the firm's investment opportunities). If it is accepted that both book and market values can be derived from objective data the ROV will be identical for all potential investors and hence can be computed using risk-neutral valuation techniques originally designed to price financial options. 

 Although data requirements are minimal, the classic school makes very strict assumptions that have limited practical relevance. Also, several scholars such as \citet{GH2001} and \citet{HH2006} present comprehensive arguments against the assumption of a complete market where it is possible to replicate a real option by its underlying asset or project. In addition, risk-neutral valuation  is designed to capture a unique value or price for all potential investors and therefore cannot capture the investor heterogeneity that is so important in practical applications. When investors are homogeneous their subjective ROV is zero and the ROV is subsumed into a unique market price \citet{M1977}.
 
Different methods for incorporating subjective information have been proposed, leading to many variants. See \citet{CA2003,HM2007,HM200702,EHH2008,CDRS2011,G2011,CKS2017,CA2019}
and others, and \citet{B2005} for a comprehensive review. These models use individual inputs, although some variables could still rely on commonly-observed prices, or rates, or other  `objective' inputs. Some subjective models are like classic models -- the assumption of perfect replicability in a complete market underpins a fundamental no-arbitrage argument which justifies the standard option pricing and hedging approach.  Others apply a subjective utility function, assuming the market is partially complete so that replication is at most imperfect.  

Unfortunately, many utility models are too complex for practical application being developed for partially complete markets which are difficult to find.\footnote{For utility models see for instance \citet{B2002, AS2004,EHH2008,CDRS2011,G2011,CKS2017}. }
Alternatively and most commonly, the literature adopts the classic risk-neutral valuation technique. For instance, \citet{CA2003} propose the marketed asset disclaimer (MAD) approach building upon a discounted cash flow (DCF) valuation model for the underlying asset. The MAD approach has become widely accepted in practice given its minimal data requirement \citep[see][and others]{BD2005,P2010}.\footnote{For other risk-neutral models see \citet{HM2007,HM200702,MPP2020}
and for practical applications see \citet{bajeux2010uncertainty,Con2018,BNRS2020,SDJ2020} and many others.}
However, the MAD approach only derives a ROV under a risk-neutral measure. 

A third strand of  research, which we  develop in this paper, applies a discounted cash-flow (DCF) analysis without requiring any form of replication. This has theoretical and practical advantages -- indeed it is the pioneering work by  \cite{CA2003} that has most often been applied in practice. We motivate our subjective model as one that uses no more data than discounted cash flow (DCF) analysis but the potential investor need not be risk-neutral. The investor has a real option to invest now, or in the future, in a project  defined by uncertain costs and revenues. The investor's views on expected future cash flows also depend on his cost of capital, and a subjective discount rate captures both the cost of capital and risk preferences. We compute the net present value of the project's actual cash flows discounted at different rates and equate the option's pay-off to their difference. The investor will perceive a non-zero option value only when he disagrees with the market about the expected return on the project.

A main advantage of the MAD approach is to connect the project value with its cash flow growth and the individual investor's required return. 
We extend this connection between project value and required returns in such as way that the subjective ROV  has economic meaning: the value of investing now, or in the future, stems from the investor disagreeing with market about the required return for the project cash flows. They could still agree with the market's projection of the cash flow amounts, but not about the required returns. The rate of return that is appropriate for the individual investor depends on the cost of capital he associates with the particular investment and his risk preferences, both of which can differ from the market view as a whole.

Our motivation follows  \cite{BM2002}, \cite{GS2010} and \cite{V2014}, all of whom assume that the investor's view of the company (or project) value can  differ from the market, arguing that different investors could perceive different subjective values. Our key idea is to incorporate this subjectivity into an investor-specific pay-off which shares the same mathematical features as an exchange or stochastic-strike option.\footnote{Since its pay-off has two stochastic components, the real option can be regarded either as an exchange option or as an option with stochastic strike. These concepts  have been applied in  real option theory before. See for instance, \citet{S2001,BBdLA2010,BMF2012,BDH2012,JdSZ2013}.
However, none of these models allows for subjective investor views and so there are important methodological  differences between our model and the previous literature. Specifically, the stochastic strike  need not be a martingale and we have no recourse to any risk-neutral measure.}
 This is appealing because it allows us to reconcile  option-pricing approaches with  traditional planning methods based on the net present value (NPV) of an investment.\footnote{Discounting the same cash flows using different discount rates implies the market is not  perfectly competitive, because not all investors operate at the same cost of capital as a market representative investor, and/or they differ in their risk preferences associated with the project's cash flows. Note that \cite{G2002} employs a general equilibrium framework to show that the ROV is zero in a perfectly competitive market.}   To compute the ROV, the early exercise boundary and the value of delay we require the NPV at any time $t$ between time 0 (when the option is valued) and time $T$ (when the investment opportunity expires). This is derived using a DCF analysis in which the NPV$_t$ is equated to the difference between the expected future cash flows discounted at two different rates, one that is subjective to the investor and the other implied by the market using the internal rate of return (IRR).\footnote{There is nothing in the theory to prevent these discount rates to be time varying (but not stochastic). However, to derive a representative forward curve of discount rates can be quite challenging in practice and for  this reason our numerical example in Excel and the code provided for all results in this article do not incorporate time-varying discount rates.} Obviously, with this definition  the NPV becomes subjective.

A  feature that distinguishes our subjective discounted cash flow (SDCF) approach from any other before is that we  define `subjectivity' in the value of a real option explicitly, arguing that it represents the profit that an investor expects from strategically exploiting what they perceive as the market's `misvaluation' of an investment. 
 This way, subjectivity stems from two inputs, the growth in expected future profits and a required return which has two elements -- the cost of capital that the investor attributes to the project and his risk preferences, the latter being a feature of the investor and this particular project.\footnote{The choice of the discount rate for the DCF valuation of projects, or equivalently the investor's required return is usually his cost of capital \citep[][]{BMM2010}. Often conventionally practitioners use the average cost of capital across their projects to discount the  future cash flows of the project in valuation, \citet{GH2001} have documented that close to 60\% of their survey respondents (from Fortune 500 companies’ financial
officers \& members of the Financial Executives Institute) takes the hurdle rate, i.e. the cost of capital specific to the project.} Taken together, these two elements define the total risk premium that the investor associates with the project -- and the subjective discount rate is the sum of a risk-free rate and this risk premium. For this total risk premium we use a subjective discount rate $r_p$ equal to the investor's expected return on the project. We further decompose risk into systematic (at least partially hedgeable) and idiosyncratic (un-hedgeable) risk factors, each having its own risk premium.  Having said this, the only two subjective inputs required for computing the real option values -- of investing now or delaying the investment  -- are the growth in expected future profits and the subjective discount rate itself, both of which are also used in the MAD model.\footnote{We ignore any differences in operational efficiency so that only the cost of capital and the investor's subjective risk preferences determine the subjective discount rate.} For this reason, our approach is very attractive for practical applications. Our discussion of the results obtained from SCDF model compared with those based on the MAD approach, using identical data inputs, exhibits some important applications for incorporating investor heterogeneity into investment valuations.

At any time $t$ between the decision time ($t=0$) and the expiry of the option  ($t=T$) a positive/negative NPV$_t$  implies an expectation at time $t$ of future cash flows that is greater/less than the cash flows that are discounted to time $t$  using the IRR.   
Then NPV$_0$ can be positive  if the investor believes the project's market value is less than it should be, and it can be negative if the investor believes the project's market value is greater than it should be.\footnote{Note that we use the term market value not market price throughout this paper because the term price only applies in a complete market where all investors agree on the same value. } If the investor has the same cost of capital and risk preferences as the market representative, we assume their subjective discount rate will be the same as the IRR, and then NPV$_0=0$.

A positive value for the real option to invest in the project arises from the investor perceiving an opportunity to maximize the NPV of cash flows throughout the life of the real option, a maximum which is only obtained if the option is exercised at the optimal time. Thus, the option value  depends on  the maximum positive discounted expected NPV$_t$ over the life of the option, or zero if all expected NPV$_t$ are zero.\footnote{Note that because NPV$_t$ could be either positive or negative at any  $t>0$,  the ROV can be positive even when NPV$_0 = 0.$} Put another way, before defining the option pay-off we  discount NPV$_t$ to time 0  for all $t$ with $0 \le t \le T$, using the investor's subjective discount rate. This way, there is no need to specify an appropriate discount rate for the pay-off itself -- a difficult choice that has been criticised by \citet{DP1994} and many others since. Indeed, by discounting all quantities to time 0 \textit{before} defining the option pay-off our model requires the same `objective' inputs as used in standard NPV calculations, which are widely accepted and are arguably the most popular approach for practical applications.\footnote{See \citet{GH2001,RR2002,HH2006,B2007,BMM2010,L2017}.}

{Using the model to simulate values for the option using different parameter values, we demonstrate several new results. We show that a non-zero ROV only arises when the investor disagrees with the market about the risk premium associated with the investment uncertainty. Similarly,  if there is a perfect hedging strategy this removes any disagreement from the investment decision making so the ROV is zero. Furthermore, we show that the value of delaying investment has a convex relationship with the volatility of any unhedged risk factors, especially when the investor has a high risk premium for the unhedged factor. These findings support the practice of isolated decision making and hedging strategies, provided there is no disagreement about risk and return expectations. By the same token, the complexity of the  decision making process can be greatly simplified when a hedging strategy is available.}

{We also generate a number of results which are supported by the approaches advocated by  \cite{HM2007,HM200702} and \cite{MW2007}. First, idiosyncratic risk is an important determinant of  the subjective ROV, even when its market risk premium is zero. This may explain an interest in taking on idiosyncratic risk (e.g. through strategies of investing in specific stocks) even when the investor is very  risk-averse. Secondly,  when the investor and the market agree on the premium of a risk that can be hedged, hedging this has almost no impact on the investor's decision making.\footnote{The market's discount rate includes a market risk premium and risk-free rate. The market risk premium can be decomposed to a sum of market risk premia of the systematic risk factors and the idiosyncratic risk. The classic asset pricing models sets the market premium for the idiosyncratic risk as zero which is the case we generate our results in.} Thirdly, the investor's perceived profitable opportunity can become slimmer when idiosyncratic risks increase; but when the idiosyncratic risk level is high, investors can rush their decisions. These risks -- and their associated risk premia -- are the main drivers of the subjective ROV. To some extent, this result also explains why the market aggregate investment level decreases with idiosyncratic risks, except when the risk level is exceptionally high, and triggers an increasing  level of investment in the market as a whole \citep{liu2021investment}. }

{The first practical advantage of our approach is its simplicity, whilst still replicating some well-known features of the more complex models already mentioned above. our model can produce intuitive results without requiring any more parameters than  traditional models -- and traditional models may also over-estimate the expected profit from an investment and produce misleading values for the opportunity to delay it. Another advantage is that our model reveals several new features about investment real options based on the obviously practical and realistic assumption that an investor has their own subjective  views about risk premia. Thirdly, our model is very general, it can be applied to any type of investment option. In contrast to growth option valuation models, which focus on the market view of the investment opportunities of a company as a whole, our model supports investment decision making for any type of investor. }

In the following: Section \ref{sec.model} describes the model; Section \ref{sec:ex}  compares numerical results with those generated using the MAD approach of \cite{CA2003} using simple examples; Section \ref{sec.simu} investigates the effects of different parameters on the value of delay; Section \ref{sec:hedging} analyses the effect of hedging;  and Section \ref{sec.concl} summarizes and concludes. All our results can be replicated by the reader. We provide an Excel workbook for the numerical  illustrations in Section \ref{sec:ex} and our code for simulations and implementations of all other results is available on request from the authors.

\section{The Model}\label{sec.model}

Consider an arbitrage-free market where an investor can acquire a project to operate and thereby produce cash flow streams. For simplicity, we assume that the project generates revenue at its maximum capacity and already exhibits an objective, profitable track record of its operating cash flows. {We do not model the project value with its own stochastic process, as in \citeauthor{H2007}, \citeyear{H2007} and \citeauthor{G2011}, \citeyear{G2011}, and others. Instead,   we model the project cash flows separately and then derive the project value accordingly, as in the DCF approach.  While some complete market models such as \citet{CA2003} are also developed under cash flow assumptions, our model is more general because we are not limited to the case of a complete market. Moreover, our setting is more intuitive because it allows for potentially negative profits. 
Furthermore, in contrast to several important works such as \citet{M1977} , \citet{I2002}, \citet{H2007} and \citet{G2011}, all of which assume that operating costs are pre-determined, we employ the more realistic assumption of stochastic operating costs which are correlated with the operating revenues. This case may arise, for instance, when project operations are likely to consume energy commodities with volatile market prices, as in \citet{S2001}.}

So the main  decision for the investor is whether to acquire the project while other real options, such as the option to rescale or suspend its production etc., are negligible.\footnote{This option is actually common in practice. For instance, 
	Barrick Gold -- as the largest gold producer in the world -- has acquired numerous high-quality mines during the past twenty years. In 1994 it took over Lac Mineral Ltd and hence acquired several large deposits including the El Indio Belt and the Veladero Project. In 1996 it acquired Aneguipa Resources Ltd. and became the sole owner of the Pierina deposit. In 2000, 2001, 2006, 2007, 2008 and 2010, Barrick Gold acquired at least 16 high-quality mines, especially in 2006 when it acquired Placer Dome Inc., adding a dozen new deposits into its global portfolio. By 2009, Barrick Gold had grown to hold the industry's largest reserves. Barrick Gold has also announced, many times, that its only operating target is to operate high-quality mines. All the mines Barrick Gold took over were identified as high-quality and all were actively producing before Barrick Gold acquired them and, once acquired, Barrick Gold kept their mining operation at full capacity, as before. In practice, rescaling or suspending the operation of an investment involves a significant cost, and the resultant change of production often affects the owning company's share prices once the information becomes public news. Therefore these decisions are commonly made in a stressed situation. However, \cite{S2001} argues that such decisions are of second order.} The projected cash flow (\textit{i.e.} the operating revenue minus the cost) from the project at time $t$ is assumed to be stochastic, and is denoted $x_t$. The real option decision is made at time 0 and it is based on projected cash flows $x_t$ at all future times $t$ with $0 \le t \le T$. We follow the standard literature on DCF models \citep[e.g.][]{CA2003} and define the project value at some future time $t$ by discounting the projected cash flows $x_{\tau}$ at all times $\tau$ with $t \le \tau\le T$. We assume the expected future cash flows $\mathbb{E}_t[x_t]$ are common to all potential investors, and that there is a market value for the project from which we can derive the IRR as the constant discount rate which equates the sum of the discounted expected cash flows  to the market value.\footnote{There are many arguments in the literature supporting this assumption. The consensus is that it is realistic for mature, stand-alone investments with publicly available financial reports on their operating efficiency, \textit{i.e.} the operating costs and revenues of their production. We make this assumption so that $\mathbb{E}_t[x_t]$ is common knowledge. There is also nothing in the model to prevent these rates from being time-varying (but non-stochastic). However, for ease of exposition in theory and because additional subjective data inputs are needed to allow for time variation, they are assumed constant throughout the life of the project.  }
	
Following the DCF literature we apply two different discount rates: the potential investor selects a discount rate $r_p$ that reflects the required return they wish to achieve from the project and this return is typically decided by (1) the investor's cost of capital which depends on their financing ability and opportunity costs and (2) the investor's risk preferences. That is, both cost of capital and risk preferences affect the risk premium the investor associates with a given investment. Similarly, the discount rate implied by the market value (i.e. the IRR, denoted $r_q$) captures a market risk premium which reflects the market's view on the average required return possible for the project and the risk preferences of a representative investor. By contrast with operational efficiency, we do not ignore differences between cost of capital and risk preferences so that a divergence between $r_p$ and $r_q$ can arise from either.

Let $\mathbb{E}_t[x_{\tau}]$ denote the expectation at time $t$ of the cash flow $x_{\tau}$ at some future time $\tau \ge t$.  For now, we assume $r_p$ and $r_q$ are constant. Now, to compare their subjective valuation of the project, $p_t$ with the observable market value at time $t$, denoted $q_t$, the  investor discounts the expected future cash flows $\mathbb{E}_t[x_{\tau}]$ to time $t$ using the two discount rates, $r_p$ and $r_q$. That is, for $0\leq t \leq \tau\leq T$ the investor compares  two different discounted cash flows, viz.:
\begin{equation}\label{eqn.subj}
p_t=\int^T_{t}\mathbb{E}_t[x_{\tau}]\exp\left(-r_{p}(\tau - t)\right)d\tau\,,
\end{equation}
and
\begin{equation}\label{eqn.equiv}
q_t=\int^T_{t}\mathbb{E}_t[x_{\tau}]\exp\left(-r_{q}(\tau - t)\right)d\tau\,.
\end{equation}
We remark that $p_{t}$ and $q_{t}$ are merely different values for the \textit{same} expected cash flows, one subjective and one implied by the project's IRR. If $p_t > q_t$ the investor associates an expected return from all future cash flows, as seen from time $t$, that is higher for himself than he attributes to the market view; and the converse applies if  $p_t < q_t$. Such divergence is possible because the investor has cost of capital, and/or risk preferences that differ from the market representative.   If the  discount rates coincide, then  $p_t = q_t$ for all $t$, $0 \le t \le T$. The investor's subjective value, $p_t$, which may be more or less than $q_t$  depends on the profit the investor expects from the investment arising from their flexibility to exploit opportunities that have been missed by the market as a whole and $p_t - q_t$ represents investor's perception of the market's possible  \textit{mis}-valuation of the project's cash flows at any future time $t$.  
 
We define the NPV of investment, as perceived by the investor at the time of the decision, as $\mbox{NPV}_0=p_0-q_0$.  This will be non-zero when the investor believes the market is over- or under-valuing the project. Now, as already proposed by \citet{CA2003} and many others, we  regard the real option as a contingent claim \textit{in addition} to the project value.  
In this case, a real option warrants the investor to exploit their maximum expected profit from the investment, not necessarily starting now but possibility starting at some point in the future. This way,  the option is written on the stream $\mbox{NPV}_t = p_t-q_t$ for all $t$ between 0, the time of the decision and $T$, the expiry time of the opportunity to invest.  Note that  NPV$_t \ne 0$  implies there is a fundamental difference between the project value represented by the market and that perceived by the investor, not only now but also in future. Although a market value $q_t$ at time $t>0$  is not realised (there is no recorded sale and purchase of the project yet) it can be implied using the IRR to discount all future  expected cash flows, not to time 0 but to  time $t$. 

Because the option allows the investor to exploit the \textit{maximum}  NPV$_t$ between 0 and $T$ the investor needs to compare a stream of  
$\mbox{NPV}_t$, for $0\le t \le T$.  That is, they make a decision at time 0, based on their judgement whether $q_t$ is a market under- or over-valuation of the project, at any time between the decision time and expiry of the option.  The real option has zero value unless at least one NPV$_t$ is positive, for some $t \in [0,T]$. To inform the real-option decision, the investor needs to capture the potential perceived mispricing by discounting the stream of $\mbox{NPV}_t$ to a single present value.  For this, they employ their own subjective rate to discount both $p_t$ and $q_t$:\footnote{If they used the IRR to discount $q_t$ instead, they would obtain the market's present value of the expected cash flow after time $t$, not their own.} 
\label{eqn.OV}
	\begin{equation} p_{t}^0=p_{t}\exp\left(-r_pt\right), \qquad 
	q_{t}^0=q_{t}\exp\left(-r_pt\right), 
	\end{equation}
and the NPV at time $t$ in time $0$ terms is:\footnote{In general,  if $t>0$ and there are cash flows between time 0 and time $t$ then $q_t^0 \ne q_0$, $p_t^0 \ne p_0$ and so also NPV$_t^0 \ne \text{NPV}_0$.}
\begin{equation}\label{eqn.NPVQ}
\mbox{NPV}_{t}^0= p_t^0-q_t^0=\mbox{NPV}_{t}\exp\left(-r_pt\right), 
 \qquad 0\leq t\leq T.
\end{equation}

Recall that the classic DCF model as in \cite{CA2003} recommends to invest when $\mbox{NPV}_0$ is positive. However, in our framework, when deciding whether to acquire the project at time 0, the investor must find $\mbox{NPV}_t$  for all $t$ with $0\le t \le T$ and would only consider acquiring this project if at least one $\mbox{NPV}_t >0$  for some $t$.\footnote{As already noted, that $\mbox{NPV}_t >0$ does not necessarily imply the investor would acquire this project, only that the project could be profitable in their view. } Set  $P_t := \mbox{NPV}_t^{0+}$, \textit{i.e.} $\mbox{NPV}_t^0$ if it is positive, and zero otherwise. Then we can define the value of the option to acquire the project as:
\begin{equation}\label{eqn.ro}
V_0 =\sup\limits_{0\leq \tau\leq T}P_\tau.
\end{equation}

This is the reward to an investor that can only be realized if they acquire the project at the right time, now or at some point in the future. Setting $v_0=V_0-P_0$ gives the value of delaying the acquisition. Specifically, if $v_0=0$ either the investor acquires the project now, or never; and if $v_0 > 0$ the investor does not acquire the project now, but might consider investing in future (depending, as always on the value of alternative investments). This view of an option to acquire extends the classic DCF theory by identifying the value of delaying the investment now for a higher perceived profit later on.\footnote{We should highlight that our construction contradicts no-arbitrage assumption made in classic risk-neutral real option valuation models \citep[such as][]{CA2003,HM2007,HM200702,MPP2020}
but there is no direct contradiction of the classic risk-neutral valuation models \textit{per se}. }

\section{Numerical Example}\label{sec:ex}
{ To demonstrate the practicality of our SDCF model, we follow a case study originally discussed in \citet{CA2003} which assumes  that the project's profit stream follows a geometric Brownian motion represented using a binomial tree. We compare the results from our model with those obtained using MAD approach, so we can assume a common set of parameter values in this exercise.  We use the MAD approach as benchmark because it is the traditional DCF model in the real-option literature \citep{BD2005, P2010} and because other risk-neutral models make completely different assumptions, typically being based on only one stochastic process for the evolution of the project value  -- see, for example, \citet{H2007} and \citet{G2011}. The MAD and SDCF  models use an identical set of inputs, and for our example we suppose:}

\begin{itemize}[noitemsep]
	\item 
	The current and future profit cash flows (in \$), $x_t$, $t = 0,1,2,\dots,T$\,.
	\item
	The growth rate, $\mu$ and volatility, $\sigma$ of this profit stream.
	\item 
	The investor's subjective discount rate $r_p$\,.
	\item 
	The current market price of the project (in \$), $q_0$\,.
\end{itemize}
To illustrate how numerical results differ between the two models we assume the following values for the inputs:
$$T=5 \mbox{~years},\quad x_0 =1,\quad \mu=20\%,\quad\sigma=30\%\quad r_p=10\%,\quad q_0=7.$$
An Excel file allows the reader to replicate the numerical example with different values of the inputs.  Both models adopt the traditional DCF framework (1) so the present value under both approaches is: $$p_0 = \sum^5_{t=0}x_0\exp((\mu-r_p)t)=\,\sum^5_{t=0}1\times\exp((20\%-10\%)\times t)=7.54.$$
Thus, $\mbox{NPV}_0 = p_0-q_0=\,7.54-7=0.54$. 

There are three main differences in the way we model the evolution of the project value in the SDCF approach. First, we build the binomial tree of the project value under a subjective measure, whereas the MAD approach adopts the risk-neutral measure. Notably, we obtain non-risk-neutral ROV without requiring any more data than the risk-neutral model. Second, the MAD approach generates simulated profit paths from which we derive dividend yields (treating profits as dividends paid from the project) and volatilities of the project value, and these are applied to construct a binomial tree for the project value. By contrast, our SDCF approach  builds a binomial tree for the profit directly and uses the relationship between profit and project value implied by the DCF framework (1) to derive another binomial tree of the project value. Third, an IRR can always be implied by the current market price but this is only used in our model, not  in the MAD approach. We use the IRR to build a binomial tree for the evolution of the future market value of the project.\\


\noindent \textit{Modelling the project value using the MAD approach}

\noindent This subsection merely applies the MAD approach explained in \cite{CA2003}, there is nothing new but it is important to include details in order to really expound on the difference between the MAD and the SDCF approach. The  MAD model requires a binomial tree of the project value which depends on the volatility of the project value and the dividend yields. To obtain these, first  we need to simulate $N=10,000$ paths of profits $x_{nt}$ -- where $n$ is the path number and $t = 1, 2, ..., 5$, assuming these follow a GBM with mean $\mu = 20\%$ and volatility $\sigma = 30\%$. The left panel of Table \ref{tab.pv} depicts the first 4 paths. The right panel of Table \ref{tab.pv} depicts the first 4 paths of the project's present value $p_{nt}$ computed as the sum of discounted future profits. For instance, for path $n=1$ at time $t=0$:
$$p_{1,0} = \sum^5_{t=0}x_{1,t}\exp(-r_pt) 
= 1\times\exp(-10\%\times 0)+0.91\times\exp(-10\%\times 1)+0.95\times\exp(-10\%\times 2)$$
$$ + 0.99\times\exp(-10\%\times 3)+1.60\times\exp(-10\%\times 4)+2.38\times\exp(-10\%\times 5)
= 5.92$$

\begin{table}[!ht]\footnotesize

	\noindent\begin{minipage}{.48\linewidth}
		\begin{center}
			\begin{tabular}{c|cccccc}
				\hline
				\multicolumn{7}{c}{Simulated paths of the profit stream, $x_{nt}$}\\\hline
				&\multicolumn{6}{c}{Year ($t$)}\\\cline{2-7}
				Path ($n$)&0&1&2&3&4&5\\\hline
				1&1&	0.91&	0.95&	0.99&	1.60&2.38\\
				2&1&	1.26&	1.66&	1.97&	2.12&	2.67\\
				3&1&	1.32&	1.72&	2.09&	2.61&	3.20\\
				4&1&	1.47&	1.67&	2.06&	2.28&	2.29\\
				\dots&\dots&\dots&\dots&\dots&\dots&\dots\\\hline
			\end{tabular}
		\end{center}
	\end{minipage}
	~\begin{minipage}{.48\linewidth}
		\begin{center} \begin{tabular}{c|c|c|c|c|c|c}\hline
				\multicolumn{7}{c}{Simulated paths of the project's present value, $p_{nt}$}\\\hline
				&\multicolumn{6}{c}{Year ($t$)}\\\cline{2-7}
				Path ($n$)&0&1&2&3&4&5\\\hline
				1&5.92&	5.42&	4.96&	4.41&	3.76&	2.38\\
				2&8.10&	7.82&	7.21&	6.10&	4.55&	2.67\\
				3&8.97&	8.77&	8.19&	7.11&	5.52&	3.20\\
				4&8.25&	7.97&	7.15&	6.03&	4.37&	2.29\\
				\dots&\dots&\dots&\dots&\dots&\dots&\dots\\\hline
			\end{tabular}
		\end{center}
	\end{minipage}
	\caption{\footnotesize The first 4 paths of the simulated profit stream (left) and the corresponding project value (right) in year 0 - 5.}\label{tab.pv}
\end{table}
Next, we take the log  returns on each path i.e.
$$R^p_{nt}=\ln\left(\frac{p_{nt}+x_{nt}}{\bar{p}_{t-1}}\right),\quad \mbox{where} \quad \bar{p}_{t-1} = \frac{1}{10,000}\sum^{10,000}_{n=1}p_{n,t-1},\quad t=1,2,\dots,5,$$ 
and we depict the first 4 paths of $R_{nt}$  in Table \ref{tab.vol}.  Here, for each column $t = 1, 2, ....,5$ we calculate the standard deviation of the 10,000 values, denoting the result $s_t$ and this is given in the penultimate row of Table \ref{tab.vol}.  Also, from the simulated paths of profit and project's value, we can easily compute the dividend yield as:
$$\delta_t = \frac{\bar{x}_{t-1}}{\bar{p}_{t-1}},\quad \bar{x}_{t-1} = \frac{1}{10,000}\sum^{10,000}_{n=1}x_{n,t-1},\quad t=1,2,\dots,5.$$
\begin{table}[!ht]\footnotesize

	\begin{center}
		\begin{tabular}{c|c|c|c|c|c}\hline
			\multicolumn{6}{c}{Simulated log changes in the project value, $R^p_{nt}$}\\\hline
			&\multicolumn{5}{c}{Year ($t$)}\\\cline{2-6}
			Path ($n$)&1&2&3&4&5\\\hline
			1&-0.27&	-0.32&	-0.34&	-0.33&	-0.26\\
			2&0.05&	0.04&	0.03&	-0.01&	-0.07\\
			3&0.15&	0.16&	0.16&	0.14&	0.12\\
			4&0.07&	0.06&	0.02&	-0.03&	-0.12\\
			\dots&\dots&\dots&\dots&\dots&\dots\\\hline\hline
			$s_t$&0.38&	0.43& 0.48& 0.53&0.57\\\hline
			$\delta_t$0.13&0.17&0.22&0.31&0.48\\\hline
		\end{tabular}
	\end{center}
	\caption{\footnotesize The first 4 paths of the log changes in the simulated project value (right) between year 0 - 5. The bottom two rows of the table depict the standard deviation of these log changes in each year, $s_t$, as well as the dividend yield $\delta_t$ computed as the ratio between the average profit in year $t-1$ and the average project value in that year.}\label{tab.vol}
\end{table}

Now with the volatility of the project value and dividend yield, we can build the binomial tree of the project value using the \citet{JR1982} parameterization, i.e. the multipliers $u$ and $d$ on upward and downward movement of the project value on the binomial tree in state $\nu$ are calculated using: 
$$\left\{\begin{array}{l}u_t = \exp\left(-0.5 s_t^2+s_t\right),\quad
d_t = \exp\left(-0.5 s_t^2-s_t\right),\\
p_{\nu u,t} = p_{\nu,t-1}(1-\delta_t)u_t,\quad p_{\nu d,t} = p_{\nu,t-1}(1-\delta_t)d_t,
\end{array}\right.$$
We assume the probability of the project value moving up at any node is $\pi = 0.5$. Note here we use a zero drift here because the MAD assumes a risk-neutral measure and because we assume, for simplicity, that the risk-free rate is zero. This means the time $t$ project value in time 0 terms, $p_t^0=p_t$.

The left panel of Table \ref{tab.binom} depicts  the resulting binomial tree of the project's value $p_t$ (or equivalently $p_t^0$). Finally at the current market price $q_0 = 7$, we compute the current and future NPV$^0$ of the project as $p_t-q_0$. We present the binomial tree of the project's NPV in the right panel of Table \ref{tab.binom}. 
For instance, $\mbox{NPV}^0_0=7.54-7=0.54$ as shown in the right panel.

\begin{table}[!ht]\footnotesize
	\begin{minipage}{.48\linewidth}
		\begin{center}
			\begin{tabular}{c|cccccc}\hline
				&\multicolumn{6}{c}{Year ($t$)}\\\cline{2-7}
				&0&1&2&3&4&5\\\hline
	&		&		&		&		&		&	29.82	\\
	&		&		&		&		&	27.62	&		\\
	&		&		&		&	20.39	&		&	14.06	\\
	&		&		&	14.36	&		&	13.02	&		\\
	&		&	10.22	&		&	9.61	&		&	5.94	\\
$p_t$	&	7.54	&		&	6.77	&		&	5.50	&		\\
	&		&	4.82	&		&	4.06	&		&	2.27	\\
	&		&		&	2.86	&		&	2.10	&		\\
	&		&		&		&	1.55	&		&	0.79	\\
	&		&		&		&		&	0.73	&		\\
	&		&		&		&		&		&	0.25	\\\hline
			\end{tabular}
		\end{center}
	\end{minipage}
	\begin{minipage}{.48\linewidth}
		\begin{center}
			\begin{tabular}{c|cccccc}\hline
	&\multicolumn{6}{c}{Year ($t$)}\\\cline{2-7}
				&0&1&2&3&4&5\\\hline
	&		&		&		&		&		&	23.59	\\
	&		&		&		&		&	21.18	&		\\
	&		&		&		&	13.70	&		&	7.34	\\
	&		&		&	7.51	&		&	6.21	&		\\
	&		&	3.26	&		&	2.70	&		&	-0.98	\\
$\mbox{NPV}_t^0$	&	0.54	&		&	-0.20	&		&	-1.46	&		\\
	&		&	-2.19	&		&	-2.93	&		&	-4.71	\\
	&		&		&	-4.14	&		&	-4.89	&		\\
	&		&		&		&	-5.45	&		&	-6.21	\\
	&		&		&		&		&	-6.27	&		\\
	&		&		&		&		&		&	-6.75	\\\hline
			\end{tabular}
		\end{center}
	\end{minipage}

	\caption{\footnotesize The left panel depicts the binomial tree for the project value with a risk-neutral drift and a 0.5 probability of the project value moving a step up or down over time. The right panel depicts the project NPV as the difference between the project value and a fixed investment price $q_0$. }\label{tab.binom}
\end{table}

\noindent \textit{Modelling the project value using the SDCF approach}\\
\noindent Our SDCF approach starts from using (2) to compute the IRR $r_q$ as the market's discount rate for the project's profits. Numerically, we back out the IRR  $r_q = 13.06\%$  from the following equation:
$$q_0=\sum^5_{t=0}x_0\exp((\mu-r_q)t) = \sum^5_{t=0}1\times\exp((20\%-r_q) t) = 7.$$

Then, instead of simulating $N$ paths of profits, we build the binomial tree of profits using the \citet{JR1982} parameterization with $\pi =0.5$ and:
$$u =\exp(\mu-0.5\sigma^2+\sigma),\quad d = \exp(\mu-0.5\sigma^2-\sigma), \quad x_{\nu u} = x_\nu u,\quad x_{\nu d} = x_\nu d,$$
where $\sigma$ is the profit volatility given as the input. The upper left panel of Table \ref{tab.rav} gives the numerical example for this tree. We then build the binomial trees of the project value and the project's market value under the DCF framework, (1) and (2). See the upper right and lower left panel of Table \ref{tab.rav}. Furthermore, to be able to compare NPV at different points in time, we further discount all future NPV to their time 0 terms, NPV$^0_t$, using the investor's discount rate $r_p$. For instance, in year 1, the project NPV$_1=p_1-q_1$ equals either $8.55-8.06=0.49$ or $5.73-5.40=0.33$. Discounting it by $r_p=10\%$, NPV$^0_1$ equals either $0.44$ or $0.30$ as shown in the lower right panel.
\begin{table}[ht!]\footnotesize
	\begin{minipage}{.48\linewidth}
		\begin{center}
			\begin{tabular}{c|cccccc}\hline
				&\multicolumn{6}{c}{Year ($t$)}\\\cline{2-7}
				&0&1&2&3&4&5\\\hline
	&		&		&		&		&		&	5.90	\\
	&		&		&		&		&	4.14	&		\\
	&		&		&		&	2.90	&		&	3.96	\\
	&		&		&	2.03	&		&	2.77	&		\\
	&		&	1.43	&		&	1.94	&		&	2.65	\\
$x_t$	&	1.00	&		&	1.36	&		&	1.86	&		\\
	&		&	0.96	&		&	1.30	&		&	1.78	\\
	&		&		&	0.91	&		&	1.25	&		\\
	&		&		&		&	0.87	&		&	1.19	\\
	&		&		&		&		&	0.84	&		\\
	&		&		&		&		&		&	0.80	\\\hline
			\end{tabular}
		\end{center}
	\end{minipage}
	\begin{minipage}{.48\linewidth}
		\begin{center}
			\begin{tabular}{c|cccccc}\hline
				&\multicolumn{6}{c}{Year ($t$)}\\\cline{2-7}
				&0&1&2&3&4&5\\\hline
					&		&		&		&		&		&	5.90	\\
	&		&		&		&		&	8.65	&		\\
	&		&		&		&	9.52	&		&	3.96	\\
	&		&		&	9.31	&		&	5.80	&		\\
	&		&	8.55	&		&	6.38	&		&	2.65	\\
$p_t$	&	7.54	&		&	6.24	&		&	3.89	&		\\
	&		&	5.73	&		&	4.28	&		&	1.78	\\
	&		&		&	4.19	&		&	2.61	&		\\
	&		&		&		&	2.87	&		&	1.19	\\
	&		&		&		&		&	1.75	&		\\
	&		&		&		&		&		&	0.80	\\\hline
			\end{tabular}
		\end{center}
	\end{minipage}
	
	~\\[12pt]
	\noindent	\begin{minipage}{.48\linewidth}
		\begin{center}
			\begin{tabular}{c|cccccc}\hline
				&\multicolumn{6}{c}{Year ($t$)}\\\cline{2-7}
				&0&1&2&3&4&5\\\hline
					&		&		&		&		&		&	5.90	\\
	&		&		&		&		&	8.53	&		\\
	&		&		&		&	9.25	&		&	3.96	\\
	&		&		&	8.92	&		&	5.72	&		\\
	&		&	8.06	&		&	6.20	&		&	2.65	\\
$q_t$	&	7.00	&		&	5.98	&		&	3.83	&		\\
	&		&	5.40	&		&	4.16	&		&	1.78	\\
	&		&		&	4.01	&		&	2.57	&		\\
	&		&		&		&	2.79	&		&	1.19	\\
	&		&		&		&		&	1.72	&		\\
	&		&		&		&		&		&	0.80	\\\hline
			\end{tabular}
		\end{center}
	\end{minipage}
	\begin{minipage}{.48\linewidth}
		\begin{center}
			\begin{tabular}{c|cccccc}\hline
				&\multicolumn{6}{c}{Year ($t$)}\\\cline{2-7}
				&0&1&2&3&4&5\\\hline
		&		&		&		&		&		&	0.00	\\
	&		&		&		&		&	0.08	&		\\
	&		&		&		&	0.20	&		&	0.00	\\
	&		&		&	0.33	&		&	0.06	&		\\
	&		&	0.44	&		&	0.14	&		&	0.00	\\
$\mbox{NPV}_t^0$	&	0.54	&		&	0.22	&		&	0.04	&		\\
	&		&	0.30	&		&	0.09	&		&	0.00	\\
	&		&		&	0.15	&		&	0.03	&		\\
	&		&		&		&	0.06	&		&	0.00	\\
	&		&		&		&		&	0.02	&		\\
	&		&		&		&		&		&	0.00	\\\hline
			\end{tabular}
		\end{center}
	\end{minipage}

	\caption{\footnotesize The upper left panel depicts the binomial tree for the project profit. The upper right and lower left panels depict the binomial trees of the project value corresponding to the profit tree from the investor's and the market's perception, $p_t$ and $q_t$, computed using \eqref{eqn.subj} and \eqref{eqn.equiv}, respectively. The lower right panel depicts the binomial tree of the project NPV in time 0 terms. It is computed as the project NPV (i.e. the difference between $p_t$ and $q_t$), discounted to the time 0 terms using the investor's discount rate (see  \ref{eqn.NPVQ}).}\label{tab.rav}
\end{table}

\noindent \textit{The real option valuation}\\
\noindent Generating the real option value from the binomial tree of the project NPV follows the same method in each approach. That is, we first compute the option pay-off $P_t=\max\left\{\mbox{NPV}_t^0,0\right\}$ and then apply  backward induction to compute the option value in each state $\nu$ as $$V_{\nu,t-1} = \max\left\{P_{\nu}, \pi  V_{\nu u,t}+(1-\pi)  V_{\nu d,t}\right\}.$$ 

Table \ref{tab.rov} presents the binomial tree of $P_t$ and $V_t$ using the MAD approach in the upper left and right panels, and those using the SDCF approach in the lower left and right panels. Comparison of the right panels shows that the real option values $V_0$ are different.  Using the MAD approach $V_0 = 0.97$  and  consequently the value of delay is $v_0 = V_0-\mbox{NPV}_0 = 1.61-0.54=1.07$. By contrast, our model generates an option value $V_0=0.54$ and therefore  delay has no value: $v_0=V_0-\mbox{NPV}_0=0.54-0.54=0$. 
\begin{table}[!ht]\footnotesize

	\noindent 
	\begin{minipage}{.48\linewidth}
		\begin{center}
			\begin{tabular}{c|cccccc}\hline
			MAD	&\multicolumn{6}{c}{Year ($t$)}\\\cline{2-7}
				&0&1&2&3&4&5\\\hline
	&		&		&		&		&		&	22.65	\\
	&		&		&		&		&	20.49	&		\\
	&		&		&		&	13.35	&		&	6.99	\\
	&		&		&	7.36	&		&	5.97	&		\\
	&		&	3.22	&		&	2.60	&		&	0.00	\\
$P_t$	&	0.54	&		&	0.00	&		&	0.00	&		\\
	&		&	0.00	&		&	0.00	&		&	0.00	\\
	&		&		&	0.00	&		&	0.00	&		\\
	&		&		&		&	0.00	&		&	0.00	\\
	&		&		&		&		&	0.00	&		\\
	&		&		&		&		&		&	0.00	\\\hline
			\end{tabular}
		\end{center}
	\end{minipage}
	\begin{minipage}{.48\linewidth}
		\begin{center}
			\begin{tabular}{c|cccccc}\hline	
            MAD	&\multicolumn{6}{c}{Year ($t$)}\\\cline{2-7}
				&0&1&2&3&4&5\\\hline
					&		&		&		&		&		&	22.65	\\
	&		&		&		&		&	20.49	&		\\
	&		&		&		&	13.35	&		&	6.99	\\
	&		&		&	7.97	&		&	5.97	&		\\
	&		&	3.68	&		&	2.99	&		&	0.00	\\
$V_0$	&	1.61	&		&	1.30	&		&	0.00	&		\\
	&		&	0.00	&		&	0.00	&		&	0.00	\\
	&		&		&	0.00	&		&	0.00	&		\\
	&		&		&		&	0.00	&		&	0.00	\\
	&		&		&		&		&	0.00	&		\\
	&		&		&		&		&		&	0.00	\\\hline
			\end{tabular}
		\end{center}
	\end{minipage}
	
	~\\[12pt]
	\begin{minipage}{.48\linewidth}
		\begin{center}
			\begin{tabular}{c|cccccc}\hline
	SDCF			&\multicolumn{6}{c}{Year ($t$)}\\\cline{2-7}
				&0&1&2&3&4&5\\\hline
&		&		&		&		&		&	0.00	\\
&		&		&		&		&	0.08	&		\\
&		&		&		&	0.20	&		&	0.00	\\
&		&		&	0.33	&		&	0.06	&		\\
&		&	0.44	&		&	0.14	&		&	0.00	\\
$P_t$	&	0.54	&		&	0.22	&		&	0.04	&		\\
&		&	0.30	&		&	0.09	&		&	0.00	\\
&		&		&	0.15	&		&	0.03	&		\\
&		&		&		&	0.06	&		&	0.00	\\
&		&		&		&		&	0.02	&		\\
&		&		&		&		&		&	0.00	\\\hline
			\end{tabular}
		\end{center}
	\end{minipage}
	\begin{minipage}{.48\linewidth}
		\begin{center}
			\begin{tabular}{c|cccccc}\hline
	SDCF			&\multicolumn{6}{c}{Year ($t$)}\\\cline{2-7}
				&0&1&2&3&4&5\\\hline
&		&		&		&		&		&	0.00	\\
&		&		&		&		&	0.08	&		\\
&		&		&		&	0.20	&		&	0.00	\\
&		&		&	0.33	&		&	0.06	&		\\
&		&	0.44	&		&	0.14	&		&	0.00	\\
$V_t$	&	0.54	&		&	0.22	&		&	0.04	&		\\
&		&	0.30	&		&	0.09	&		&	0.00	\\
&		&		&	0.15	&		&	0.03	&		\\
&		&		&		&	0.06	&		&	0.00	\\
&		&		&		&		&	0.02	&		\\
&		&		&		&		&		&	0.00	\\\hline
			\end{tabular}
		\end{center}
	\end{minipage}
	\caption{\footnotesize The upper left and right panels depict the pay-off and the value of the option to invest in the project, respectively, using \citet{CA2003}'s approach. By contrast, the lower left and right panels depict the pay-off and the value of the option to invest in the project using our subjective discounted cash flows (SDCF) approach.}\label{tab.rov}
\end{table}

The SDCF approach always treats the valuation of the project as the sum of discounted expected profits in the future. So when the profit at any point in time moves up, both the investor's project value $p_t$ and the market value $q_t$ will move up. By contrast, a project value $p_t$ in the MAD method is compared with a  fixed market value $q_0$, so a rise in $p_t$ leads to a greater NPV$_t$, compared with our approach. As a result the dispersion of the NPV process is smaller in our model. Therefore, our real option value is also smaller, as is the value of delay, compared with the MAD approach. Note that the MAD approach could depart from a fixed market price assumption and in that case one might expect a different ordering in ROVs, based on the same inputs.

\section{Cash Flow Models}\label{sec:GBM}
 In the previous section the binomial tree models of profit evolution were used for a practical illustration. To further investigate our model implication on the subjective ROV and the input parameters requires a more sophisticated methodology, and here we consider  a standard two-factor continuous-time model for revenues and costs. This has been used for ROV by several authors, such as \cite{S2001}, \citet{SLC2011}, \cite{BMF2012}, and others. 	A more elaborate model generates more detailed output, and in this case we can use it  to generate  early exercise boundaries when it is optimal to delay investment, as well as calculating real option values. However, by the same token, a more elaborate model requires more sophistication for its resolution. We adopt \citet{LS2001}'s Least-Square Monte Carlo (LSM) approach to simulate the real option value. This means that our numerical results are computed in  a discrete-time framework.

	 We generate stochastic costs and revenues and hence derive the cash flows as  $x_{t}=x_{1t}-x_{2t}$. That is, we assume that the project's operation generates a stream of revenues $x_{1t}\,$, associated with a series of costs $x_{2t}\,$, with evolutions represented by correlated geometric Brownian motions (GBMs):
\begin{align}\label{eqn.xcZB}
\frac{dx_{it}}{x_{it}} = \mu_idt+\sigma_id\mathcal{Z}_{it}\,,\quad
\mathcal{Z}_{it}=\rho_{it}~ \mathcal{B}_{0t}+\sqrt{1-{\rho_{i}}^2}~\mathcal{B}_{it}\,,\quad i=1,2,\quad 0\leq t\leq T\,,
\end{align}
where $\mu_{i}$ and $\sigma_{i}$ denote the expected growth rate and the volatility of  $x_{it}\,$; $\mathcal{B}_{jt}\,$, $j=0,1,2\,$, are three orthogonal Brownian motions; and $\rho_{i}$ is the correlation between $\mathcal{Z}_{it}$ and $\mathcal{B}_{0t}$, so that the correlation between costs and revenues is $\rho =\rho_{1}\rho_{2}$. Now,  for all $\tau \in [0,T]$,
$$\mathbb{E}_0[x_{\tau}]= x_{10}\exp\,(\mu_1\tau) - x_{20}\exp\,(\mu_2\tau)$$ and we use \eqref{eqn.equiv} for $t=0$ to compute $r_q$. But we do not need to compute expected cash flows $\mathbb{E}_t[x_\tau]$ at all, because we can derive the NPV$_t^0$ directly by simplifying \eqref{eqn.subj} and \eqref{eqn.equiv} when revenues and costs are correlated GBMs, \textit{i.e.}
\begin{equation}\label{eqn.po}
{\mbox{NPV}_t^0} = c_{1t}\, x_{1t} - c_{2t}\, x_{2t}\,,
\end{equation}
	where $c_{it}$, $i=1,2$, is a deterministic function of $t$\,:
	\begin{align}\label{eqn.c}
	c_{it}=	\left[\frac{\exp((\mu_{i}-r_p)(T-t))-1}{\mu_{i}-r_p}-\frac{\exp((\mu_{i}-r_q)(T-t))-1}{\mu_{i}-r_q}\right]\exp(-r_pt).
	\end{align}
See Appendix \ref{app.npv} for the derivation of \eqref{eqn.po} and \eqref{eqn.c}. This formulation shows that ${\mbox{NPV}_t^0}$ is just a  difference between scaled revenues and costs, and it greatly simplifies the simulation of ${\mbox{NPV}_t^0}$ for computing the real option value when it does not have a closed-form solution.

The inputs for LSM are: some initial values   $x_{10}$ and $x_{20}$, the parameters $\mu_i$ and $\sigma_i$ $(i = 1,2)$ which are used to simulate future revenues and costs, and assumed values for $r_p$ and $q_0$.  First we compute $r_q$ using \eqref{eqn.equiv} and generate $n$ paths for $x_{it}$, between time $0$ and $T$. Then we compute $n$ paths of the project $\mbox{NPV}^0$ from the simulated $x_{it}$ paths using \eqref{eqn.po}. The left panel in Table \ref{tab.lsm} illustrates an example of a million simulated $\mbox{NPV}^0$ paths, between time 0 and $T$ with  $T$ set at $4$. Next we use LSM  to generate the optimal option pay-offs for all ${\mbox{NPV}_t^0}$ paths between time 0 and $T$. The right panel of Table \ref{tab.lsm} tabulates  the optimal option pay-offs computed for the ${\mbox{NPV}_t^0}$ paths that are shown on the left.  The real option value $V_0$ is then the average of all possible optimal pay-offs from time $t=0$ to $4$, \textit{i.e.} $(10.98+8.75+\dots+20.06+\dots+2.44+\dots)/10^6=19.57$ million dollars.

			\begin{table}[!ht]\scriptsize
\centering
				\begin{minipage}[t]{.48\linewidth}\vspace{0pt}\centering
			\begin{tabular}{lrrrrr}
				\multicolumn{6}{c}{Simulated ${\mbox{NPV}_t^0}$ paths}\\
				\toprule
				\multicolumn{1}{l}{Path}&\multicolumn{1}{c}{$t=0$}&\multicolumn{1}{c}{$t=1$}&\multicolumn{1}{c}{$t=2$}&\multicolumn{1}{c}{$t=3$}&\multicolumn{1}{c}{$t=4$}\\
				$1$& 15.94& 	-12.07& 	 10.98& 	-10.58& 	-2.38\\ 
				$2$& 15.94& 	-21.42& 	-20.11& 	 8.75& 	 3.39\\ 
				$\vdots$&$\vdots$&$\vdots$&$\vdots$&$\vdots$&$\vdots$\\
				$9$& 15.94& 	 20.06& 	-0.62& 	7.69& 	-3.38\\ 
				$\vdots$&$\vdots$&$\vdots$&$\vdots$&$\vdots$&$\vdots$\\
				$16$&15.94&	3.26&	-1.59&	0.39&	2.44\\
				$\vdots$&$\vdots$&$\vdots$&$\vdots$&$\vdots$&$\vdots$\\
				$92$&15.94&	-10.44&	-0.44&	0.28&	-1.37\\
				$\vdots$&$\vdots$&$\vdots$&$\vdots$&$\vdots$&$\vdots$\\
				$10^6$& 15.94 	&-20.35 	&-12.46 	&-4.58 	 &0.01 \\
				\bottomrule
			\end{tabular}
		\end{minipage}
\begin{minipage}[t]{.48\linewidth}\vspace{0pt}\centering
			\begin{tabular}{lrrrrr}
				\multicolumn{6}{c}{Optimal option pay-off for all ${\mbox{NPV}_t^0}$ paths}\\
				\toprule
				\multicolumn{1}{l}{Path}&\multicolumn{1}{c}{$t=0$}&\multicolumn{1}{c}{$t=1$}&\multicolumn{1}{c}{$t=2$}&\multicolumn{1}{c}{$t=3$}&\multicolumn{1}{c}{$t=4$}\\
				$1$&0.00&0.00&	\color{red} 10.98& 0.00& 0.00\\
				$2$&0.00&0.00&0.00&	\color{red} 8.75&0.00\\
				$\vdots$&$\vdots$&$\vdots$&$\vdots$&\bf $\vdots$&\bf $\vdots$\\
				$9$&0.00	&\color{red} 20.06&	0.00	&0.00	&0.00\\
				$\vdots$&$\vdots$&$\vdots$&$\vdots$&\bf $\vdots$&\bf $\vdots$\\
				$16$&0.00&	0.00&	0.00&0.00&\color{red} 	2.44\\
				$\vdots$&$\vdots$&$\vdots$&$\vdots$&\bf $\vdots$&\bf $\vdots$\\
				$92$&0.00&	0.00&0.00& 0.00& 0.00\\
				$\vdots$&$\vdots$&$\vdots$&$\vdots$&$\vdots$&$\vdots$\\
				$10^6$&0.00&	0.00&	0.00&	0.00	&0.00\\
				\bottomrule
			\end{tabular}
		\end{minipage}
					\caption{Numerical illustration of computing the ROV with simulated project NPVs.\\ \scriptsize{All values in the tables are in \$ million. LSM  computes the optimal option pay-off at a given time on a given path backwards, i.e. from time $t=4$ to $t=0$ in our example. We first cross out the paths with zero real option pay-offs at both time $t=3$ and $4$. Only using the remaining paths we estimate a quadratic regression of $P_4$ on ${\mbox{NPV}_3}^0$, resulting in the conditional expectation function  $\mathbb{E}[P_4|{\mbox{NPV}_3}^0] = \hat{\alpha}_3+\hat{\beta}_{3,1}x_3+\hat{\beta}_{3,2}x_3^2$. 
							Then we make the exercise decision at $t=3$ based on the maximum of $P_3$ and $\mathbb{E}[P_4|{\mbox{NPV}_3}^0]$, \textit{i.e.} the expected pay-off to delay the exercise to time $t=4$. For each path at $t=3$, if the decision is not to exercise, the option pay-off is now set to 0; otherwise, the option pay-off at $t=4$ is set to 0. We also put back the paths that we crossed out before the regression. We repeat these steps for time $t=2$ and $3$, time $t=1$ and $2$, and time $t=0$ and $1$, thus achieve the optimal pay-offs for all $\mbox{NPV}^0$ paths between time $t=0$ and $4$.  }}\label{tab.lsm}
	\end{table}

\begin{table}[!ht]\scriptsize
\centering
	
	\begin{minipage}[t]{.35\linewidth}\vspace{0pt}\centering
		\begin{tabular}{ccc}
			\multicolumn{3}{c}{(a) Early exercise boundary}\\
			\toprule
			$t$&${\mbox{NPV}_{t,U}}^0$&${\mbox{NPV}_{t,L}}^0$\\
			\midrule
			0&15.94&0\\
			1&318.99&10.21\\
			2&456.42&4.46\\
			3&734.11&1.14\\
			\bottomrule
		\end{tabular}
	\end{minipage}
	\begin{minipage}[t]{.55\linewidth}\vspace{0pt}\centering
		(b) The lower early exercise boundary of the exercise of our option to acquire the project, and paths 1, 2, 9, 16, 92, 10$^6$ of ${\mbox{NPV}_t^0}$.\\\vspace{-2.5cm}
		\includegraphics[width=.8\textwidth]{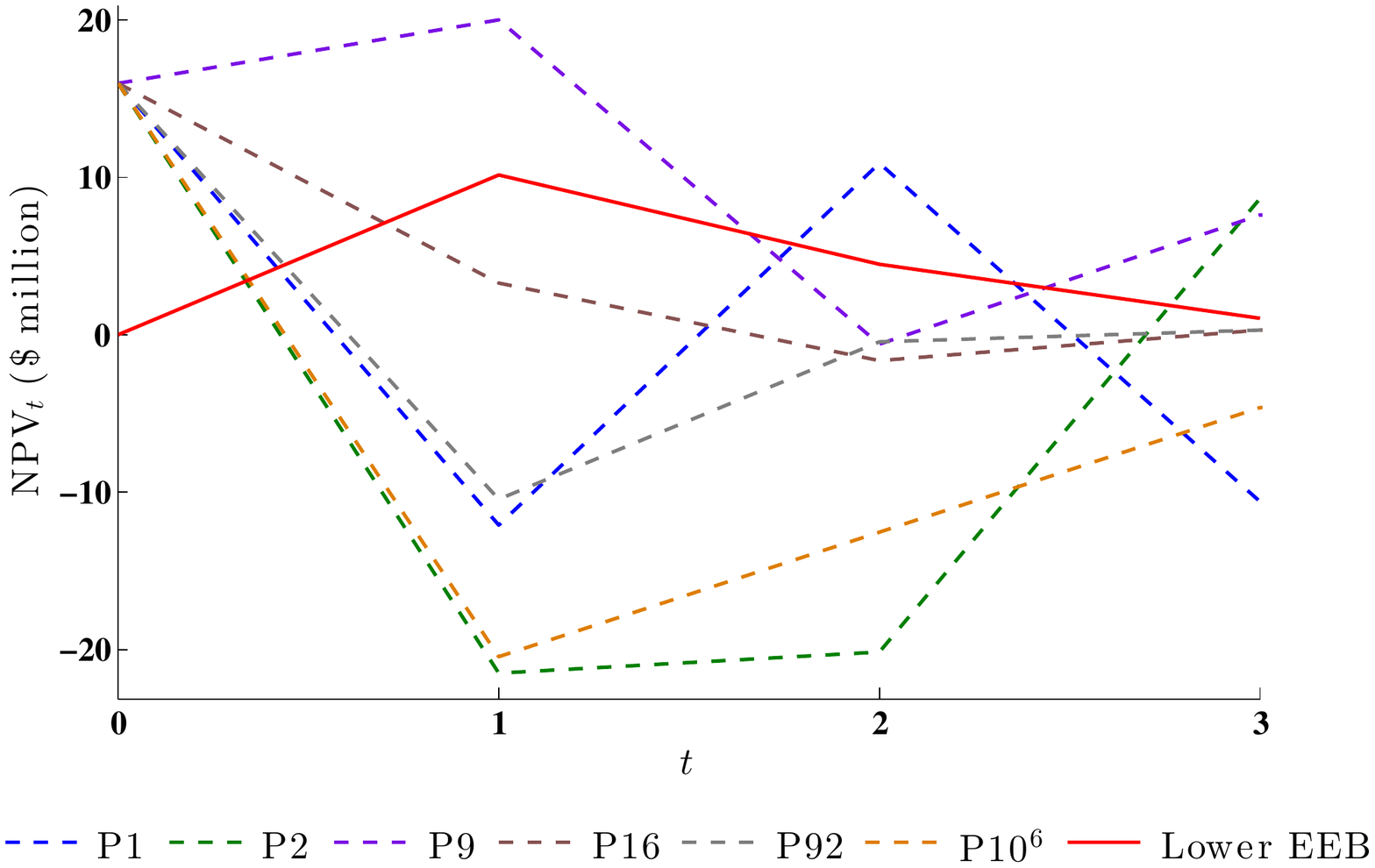}\vspace{-2.5cm}
	\end{minipage}
	\caption{Early early boundary values computed using the LSM approach.
		\\ 
		\scriptsize{Mathematically, when ${\mbox{NPV}_t^0}$ is on the boundary, $P_t=\mathbb{E}[P_{t+1}|{\mbox{NPV}_t}^0]$, the time $t$ expected option pay-off at time $t+1$, and so, the investor would be indifferent with exercising the option at time $t$ or $t+1$. 
				The LSM approach runs a quadratic regression of ${\mbox{NPV}_t^0}$ on $P_{t+1}$ and results in the estimation of $\mathbb{E}[P_{t+1}|{\mbox{NPV}_t}^0]$:
		$\mathbb{E}[P_{t+1}|{\mbox{NPV}_t}^0] = \hat{\alpha}_t+\hat{\beta}_{t,1}{\mbox{NPV}_t}^0+\hat{\beta}_{t,2}\left({\mbox{NPV}_t}^0\right)^2.$
				The estimates of the regression coefficients for our numerical example is reported in the left panel above. 
				Solving $P_t=\mathbb{E}[P_{t+1}|{\mbox{NPV}_t}^0]$ for the boundary values of $\mbox{NPV}_t^0$ is therefore equivalent to finding the non-negative roots of the following function:	$\hat{\alpha}_t+(\hat{\beta}_{t,1}-1){\mbox{NPV}_t}^0+\hat{\beta}_{t,2}\left({\mbox{NPV}_t}^0\right)^2=0.$
		}} \label{tab.eeb}
\end{table}

The advantage of adopting the LSM approach to simulate the real option value is that it also allows us to compute the early exercise boundaries of the real option, \textit{i.e.} the upper- and lower- boundary values of ${\mbox{NPV}_t^0}$ between which the real option will be early exercised. 
The left panel of Table \ref{tab.eeb} summarises the early exercise boundaries for time $t=0,1,2,3$.
Also see the right panel for a graphic illustration of the lower early exercise boundary and paths $1$, $2$, $9$, $16$, $92$ and $10^6$; the upper boundary for $t=1,\,2,\,3$ is too high for the $\mbox{NPV}_t$ to reach and hence it is not depicted in the figure. At any time $t > 0$, when a path yields a positive ${\mbox{NPV}_t^0}$ lying between the early exercise boundaries, it would exceed $\mathbb{E}[P_{t+1}|{\mbox{NPV}_t}^0]\,$ and hence the investor would exercise the option immediately instead of deferring it. For instance, on path P1 (blue) the investor defers at $t=1$ and exercises at $t=2$. 

\section{Value of Delay}\label{sec.simu}
\titleformat{\subsection}[block]{\itshape}{\normalfont\bfseries Observation \thesubsection}{1em}{}

Now we turn to investigate the effect of different parameters on the value of delay $v_0$ using the SDCF approach with the correlated GBM cost and revenue assumptions \eqref{eqn.xcZB} adopted in Section \ref{sec:GBM}. This section provides two observations we obtain as our sense checks. First we demonstrate that the SDCF approach captures the subjective ROV arising from the investor's perceived misvaluation of the market on the investment opportunity attached to the project. We achieve so by providing the limiting case of a zero subjective ROV when there is no perceived misvaluation, \textit{i.e.} the investor does not expect any return above the market rate through exercising the option of invest. This limiting case distinguishes the SDCF approach with the existing models which are theoretically ill-defined and thus fail to draw the case. Second, we show in this section, that the subjective ROV increases with the future uncertainty in the project value which has been widely acknowledged in the literature.
 
We set the following base case:
\begin{align}\label{eqn.initial}
T=5, \quad x_{1,0}=x_{2,0} = 5\quad\mbox{and}\quad\mu_1=\mu_2 = \sigma_1 = \sigma_2=\rho=r_p=r_q=30\%\,.
\end{align} 
so that the project value $p_0$ and $q_0$, $\mbox{NPV}_0$ and the value of delay $v_0$ are all zero. {This way, an investor will be indifferent between investing now or not, under both the traditional DCF approach and our SDCF. There is no particular reason for choosing these values. Nor is there any advantage in doing so. We selected values with means, correlation and volatilities al being equal merely for simplicity. However, readers may use our Python code to change any of these values, and re-run  simulations accordingly.}

From here we can observe the effect of a value change in any parameter on the option value in isolation. Also the revenue and cost processes are set to have the same distribution in the base case; this way, the effect of changes in  cost and revenue on the value of delay are of comparable magnitude.

\subsection{{\normalfont\bf The disagreement effect.}
	The value of delay stems from a difference in discount rates, $r_p \ne r_q$. }\label{sec.disagree}

\begin{figure}[!ht]\centering
	\begin{minipage}{.45\linewidth}\centering
		\includegraphics[width=1\linewidth]{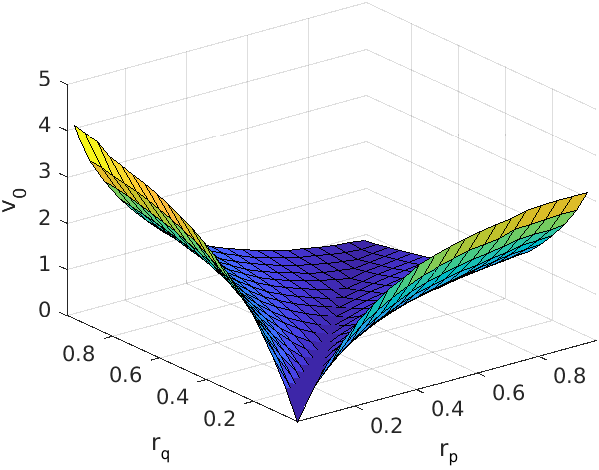}
	\end{minipage}\qquad
	\begin{minipage}{.45\linewidth}\centering
		\includegraphics[width=1\linewidth]{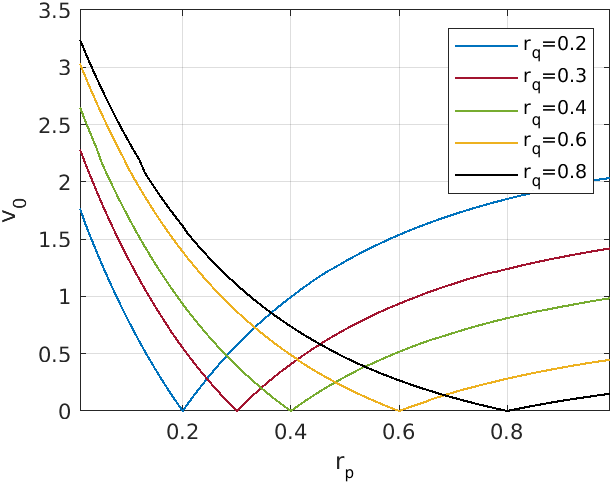}
	\end{minipage}
	\caption{\footnotesize The  ROV  for different values of the investor and market discount rates. 
		The left panel takes the discount rates each between $1\%$ and $99\%$. The right panel shows cross-sections of the  ROV  surface on the left. The blue, red, green, yellow and black solid lines are the sections drawn at the investor rate $r_p=20\%$, $30\%$, $40\%$, $60\%$, and $80\%$.  Other parameters set at their base values in \eqref{eqn.initial}. }
	\label{fig.RR}
\end{figure}

First we analyse the relationship between the difference in discount rates $r_p - r_q$ and  the value of delay. Then we show that the size of NPV$_0$ mitigates this effect so that delay is more likely to have a positive value when NPV$_0$ is small. Initially, we allow only discount rates to vary from our base case \eqref{eqn.initial}, holding the project value $q_0$ at zero. In this case NPV$_0 =0$ and so $v_0 = V_0$, i.e. the value of delay is the same as that of the real option to invest now. Figure \ref{fig.RR} plots the  option values  as a function of $r_p$ and $r_q$. This clearly shows that this option value is not necessarily driven by the discount rates \textit{per se} but by their \textit{absolute} difference: the project's net cash flows can deviate from zero in the future and any difference between the discount rates, positive or negative, increases the dispersion of future net cash flows in time 0 terms, which in turn, increases the real option value. This is consistent with our argument that the value of the option to invest in the project arises from the investor's perceived misvaluation of the project in the market. The greater this perceived misvaluation, the higher the value of the option to invest in the project.

We highlight this observation because it demonstrates the ability to distinguish between two potential reasons for a zero NPV at time $0$. It can simply be that $p_0=q_0=0$, which is consistent with the consensus of the literature  since \citet{MS1986} that allows an investor to reconsider the decision later, because NPV$_t$ could be positive for some $t \in (0,T]$. Alternatively, it could be that NPV$=0$ because the investor agrees with the market on the discount rate, $r_p = r_q$. In this case, the SDCF approach generates a zero option value exactly as we argued before: when the investor has the same risk premium as the market the NPV is zero throughout the project's life so the option to acquire or defer is valueless at any point in time.\footnote{ Although the  project values $p_t$ and $q_t$ need not be zero, if $r_p=r_q$ then he  can invest at any time but the investor will expect no more than the market's perceived return on the investment, so NPV$_t = 0$ for all $t \in [0,T]$} We shall see later that, irrespective of any other model parameters, whenever $r_p=r_q$ the option to delay has no value.\footnote{Instead of a stochastic market value, numerous existing models assume a fixed price for the project  \citep[\textit{e.g.}][]{CA2003,CDPVN2011,CDRS2011,Tou2013,HGSR2017,Con2018}
in which case NPV$_t$ always has a chance, however small, to be positive. Thus, there is always a positive value in delay. By contrast, in our SDCF approach the market value is derived from uncertain cash flows discounted at rate $r_q$, and when $r_q=r_p$, there is never a positive value to delay.}

\begin{figure}[!ht]\centering
	\begin{minipage}{.24\linewidth}\centering
		\footnotesize (a) $x_{20}=4.5$\vspace{6pt}
		\includegraphics[width=1\linewidth]{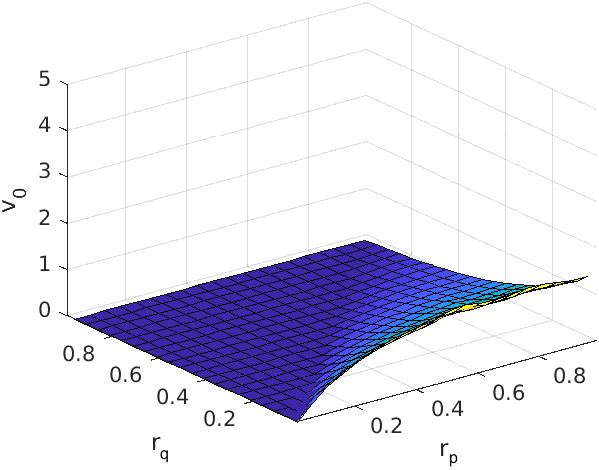}\vspace{3pt}
		\includegraphics[width=1\linewidth]{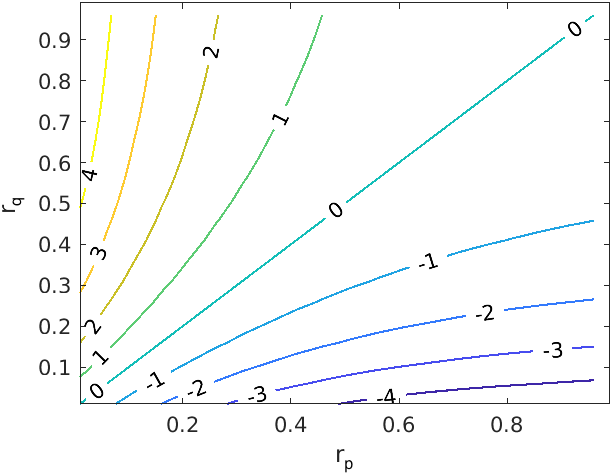}
	\end{minipage}
	\begin{minipage}{.24\linewidth}\centering
		\footnotesize(b) $x_{20}=4.8$\vspace{3pt}		
	\includegraphics[width=1\linewidth]{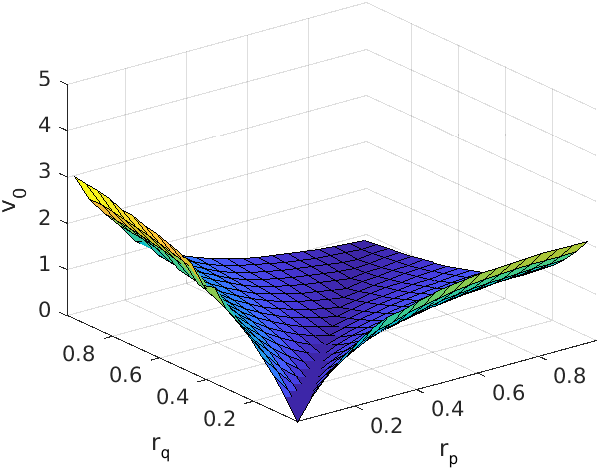}\vspace{6pt}
	\includegraphics[width=1\linewidth]{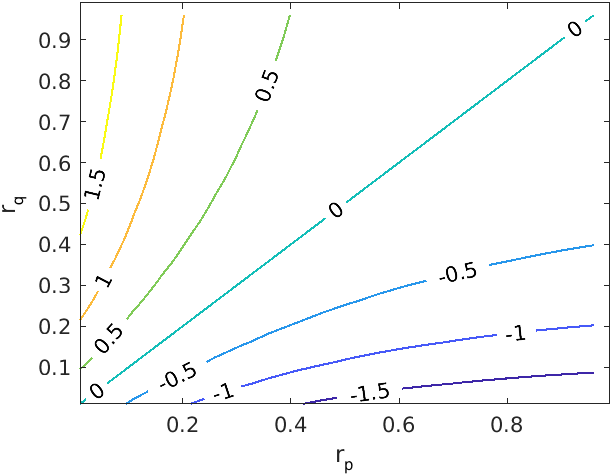}
	\end{minipage}
	\begin{minipage}{.24\linewidth}\centering
		\footnotesize(c) $x_{20}=5.2$\vspace{3pt}		
	\includegraphics[width=1\linewidth]{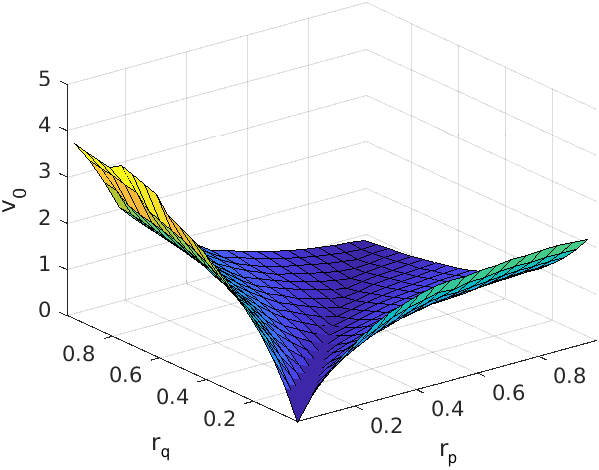}\vspace{6pt}
	\includegraphics[width=1\linewidth]{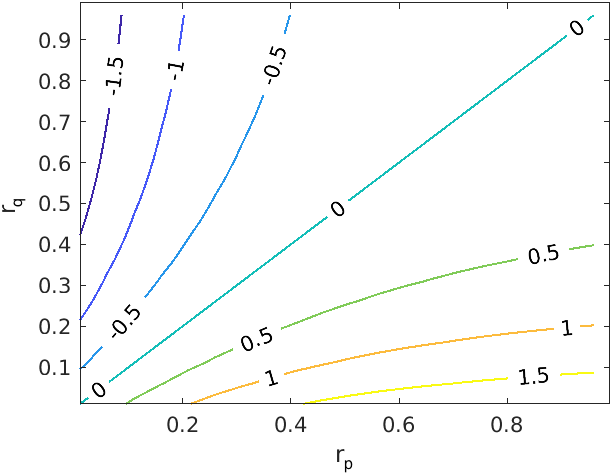}
	\end{minipage}
	\begin{minipage}{.24\linewidth}\centering
		\footnotesize(d) $x_{20}=5.5$\vspace{3pt}
	\includegraphics[width=1\linewidth]{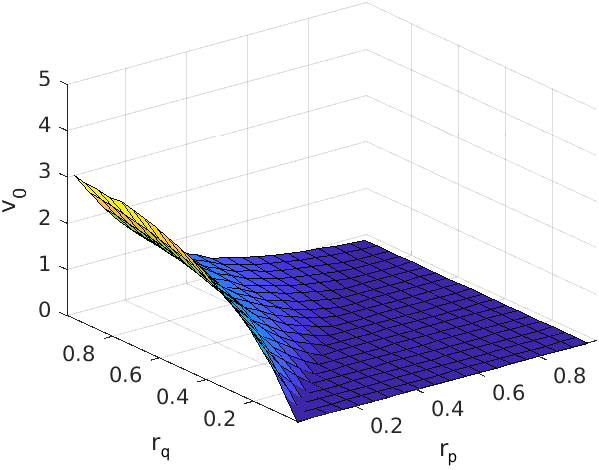}\vspace{6pt}	\includegraphics[width=1\linewidth]{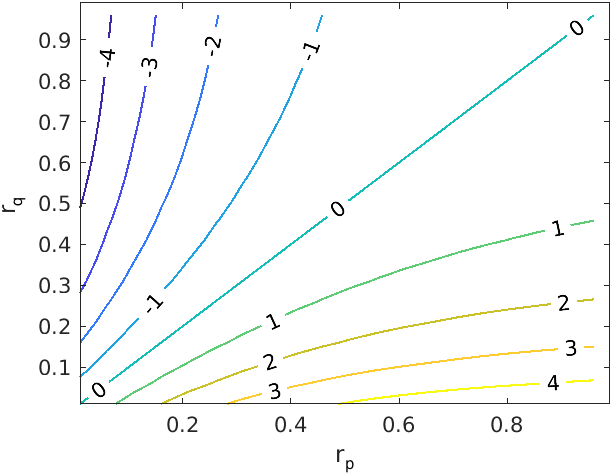}
	\end{minipage}
	\caption{\footnotesize The value of delaying the investment (first row of panels) and the NPV (second row) for different values of the investor and market discount rates, with the current cost at 4.5 (first column), 4.8 (second column), 5.2 (third column) and 5.5 (last column). Each discount rate varies between $1\%$ and $99\%$. Other parameters set at their base values in \eqref{eqn.initial}. 
	}
	\label{fig.RRnpv<>0}
\end{figure}

Next we examine the disagreement effect when NPV$_0 \ne 0$, which is perhaps more common in practice. 
Recall Figure \ref{fig.RR}, with NPV$_0 = 0$, where the value of delay arises when the investor requires a rate of return $r_p$ which differs from the market implied IRR, $r_q$.  Now, Figure \ref{fig.RRnpv<>0} shows that the value of delay decreases as the absolute value of NPV$_0$ increases. With an initial revenue, $x_{10}=5$, the four columns of panels correspond to an initial cost, $x_{20}= 4.5$, 4.8, 5.2 and 5.5 from left to right, i.e. the initial profit, x$_0 = 0.5, 0.2, -0.2,$ and  $-0.5$. The first row depicts the value of delay and it is clear that this is smaller when $|x_0| = 0.5$ than when $|x_0| = 0.2$, i.e. for the outer two columns.  This is not surprising -- like any American option, the higher  the option pay-off,  the lower  the option premium in addition to the option pay-off. The second row shows level sets of the NPV$_0$ for different pairs of discount rate $r_p$ and $r_q$. Although the value of delay in Figure \ref{fig.RR} looks roughly symmetric in $|r_p - r_q|$ it is no longer the case in Figure  \ref{fig.RRnpv<>0}. Comparing rows we see that the  value of delay can be 0 when the NPV is positive. Thus, the mitigating effect that NPV has on the value of delay is much greater when it is positive.

\begin{figure}[!ht]\centering
	\begin{minipage}{.45\linewidth}\centering
		\includegraphics[width=1\linewidth]{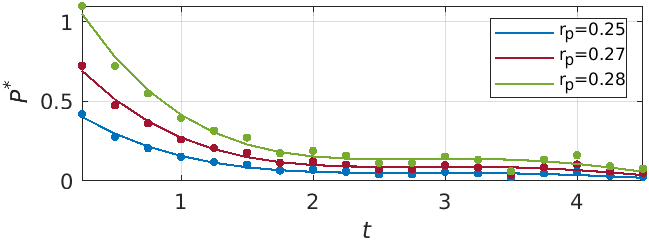}\vspace{6pt}
		\includegraphics[width=1\linewidth]{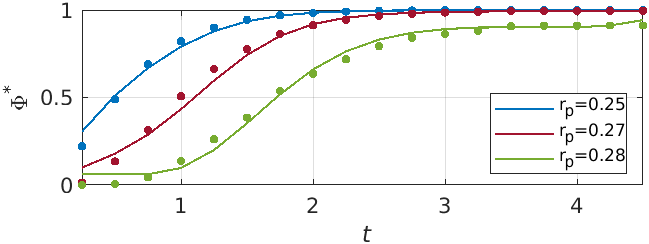}
	\end{minipage}\quad
	\begin{minipage}{.45\linewidth}\centering
		\includegraphics[width=1\linewidth]{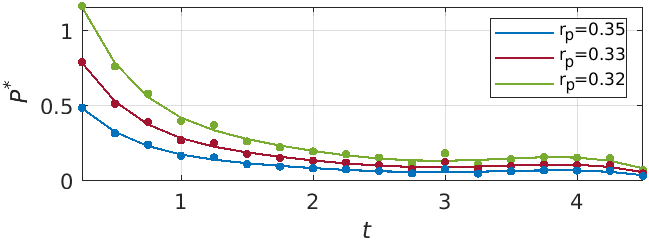}\vspace{6pt}
		\includegraphics[width=1\linewidth]{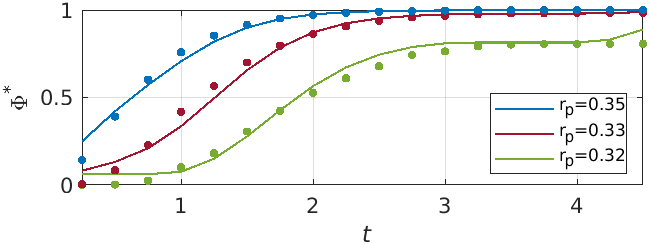}
	\end{minipage}
	\caption{\footnotesize The  lower exercise boundary $P^*$ and the probability of early exercise $\Phi^*$ with base parameters \eqref{eqn.initial} and $r_q=30\%$ in particular. On the left, $r_p<r_q$ with  $r_p=25\%,\,27\%,\,28\%$ and on the right $r_p>r_q$ with $r_p=,32\%,\,33\%,\,35\%$. The dots are actual boundary and probability values whilst the lines are fitted polynomial function values.}\label{fig.EEBrs}
\end{figure}

The early exercise boundary is the lower bound $P^*$ on the option pay-off which indicates it is optimal to invest. We also calculate $\Phi^*=\mbox{Pr}(P\geq P^*)$ i.e. the cumulative probability of an early exercise throughout the option's life. For each set of parameter values, we generate $P^*$ using the LSM approach to simulate 10,000 paths of the project values from time 0 to $T$. Then at each point in time $t$, we estimate $\Phi^*$ by  counting the number of paths on which the option will be exercised at or before time $t$ and dividing by 10,000. The disagreement effect reduces the lower early exercise boundary and drives up the cumulative probability of an early exercise, as displayed in Figure \ref{fig.EEBrs}. The greater the absolute difference between the investor's risk premium and the market risk premium (held constant here at 30\%), the more likely are they to invest at an earlier stage. This is because near-term profits is relatively more important than long-term profits due to discounting over time. This intuition supports our novel interpretation: the investors perception of the market's mispricing of the project increases with divergence of the two discount rates, which is greater in the short-term than the longer term.

\subsection{{\normalfont\bf The volatility effect}\quad The value of delay increases with uncertainty in future project values. The greater the risk-aversion or cost of capital of the investor, relative to the market, the less sensitive the value of delay to an increase in the volatility of costs.}\label{sec.vol}

\begin{figure}[!ht]
	\centering
	\begin{minipage}{.32\linewidth}\centering
		\footnotesize (a) $\sigma_2=0$\vspace{6pt}
\includegraphics[width=1\textwidth]{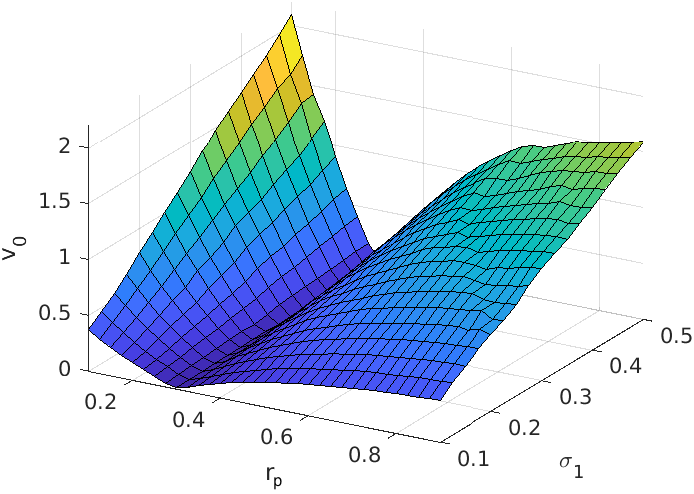}
	\end{minipage}
	\begin{minipage}{.32\linewidth}\centering
		\footnotesize (b) $r_p<r_q$\vspace{6pt}
	\includegraphics[width=1\textwidth]{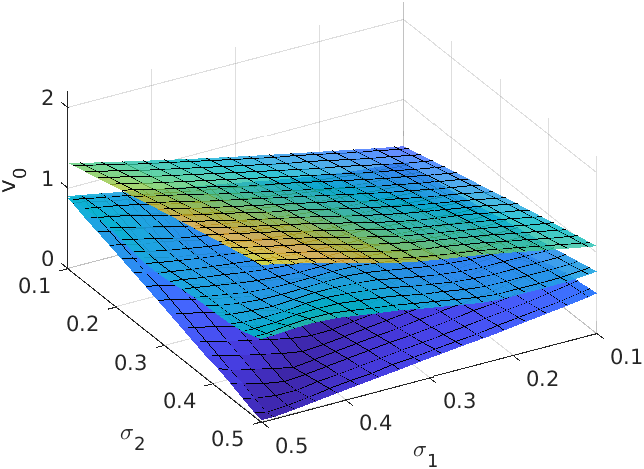}
\end{minipage}
	\begin{minipage}{.32\linewidth}\centering
		\footnotesize (c) $r_p>r_q$\vspace{6pt}
	\includegraphics[width=1\textwidth]{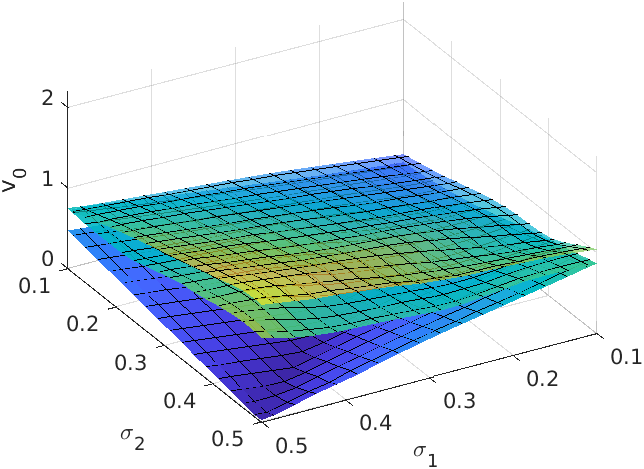}
\end{minipage}
	\caption{\footnotesize The value of delay, $v_0$  given a range for revenue and cost volatilities, i.e. $\sigma_1$ and $\sigma_2$  in \eqref{eqn.xcZB} and discount rates $r_p$ with other parameters set at their base values in \eqref{eqn.initial}.  Panel (a) on the left depicts $v_0$ under the deterministic cost assumption, i.e. $\sigma_2 = 0$. Here $\sigma_1$ varies  between $5\%$ and $50\%$ and $r_p$ varies between $20\%$ and $90\%$. The middle and right panel depicts $v_0$ as  $\sigma_1$ or $\sigma_2$ varies between $0.10$ and $0.35$. In case (b) $r_p =20\%, r_q=30\%$ and in case (c) $r_p=30\%, r_q=20\%$. On these two plots the three surfaces correspond to three different revenue-cost correlations, $\rho=-1$ (top), $\rho=0$ (middle), $\rho = -1$ (bottom). 
	 }\label{fig.volSf}
\end{figure}

In Figure \ref{fig.volSf} we make the base parameter value assumption \eqref{eqn.initial} so that $r_q=30\%$, NPV$_0=0$ and $p_0=q_0=v_0=V_0$. We consider the effect of uncertainty in profits on the value to delay. We ask whether it is always the case that increasing either cost or revenue volatilities enhances the value of delay.  To this end, Figure \ref{fig.volSf} depicts three scenarios. On the left, panel (a) assumes costs are deterministic $(\sigma_2=0)$ and plots $v_0$ as a function of $r_p$ and the revenue volatility $\sigma_1$. When $r_p=30\%$ we have $r_p=r_q$ and our observations 5.1 continue to hold. Now we allow $r_p \ne r_q$ and a slice through the surface (a) at any value for $\sigma_1$ yields a familiar shape, i.e. it is the absolute difference between $r_p$ and $r_q$ which matters. The point to note here, though, is that $v_0$ increases with $\sigma_1$ at any value $r_p \ne r_q$. This is because the option protects the investor from downside risk in revenues; the higher the revenue volatility, the higher its downside risk and hence  both $V_0$ and $v_0$ increase.\footnote{Similar remarks hold when $\sigma_1 = 0$ and we allow $\sigma_2$ to vary except that now the option can protect the investor from the upside risk in costs and gives him a greater chance to profit from exceptionally low cost. Both of these volatility effects are consistent with the general belief that the more volatile is the future project value, the more valuable  the option to invest \citep[\textit{e.g.} see][and many others]{CA2003,AS2004,BFJLR2008,CDRS2011,MCCC2011,F2015,MCP2018}: both the volatilities of the revenue and cost can drive the project value dispersion. Moreover, our observation on the cost volatility effect adds to the conclusions drawn from a literature that assumes constant costs.} 

Figures \ref{fig.volSf}{\color{red}(b)} and \ref{fig.volSf}{\color{red}(c)} both depict three surfaces representing the value of delay for three different values of $\rho$, the correlation between revenues and costs. Note that $v_0$ increases as $\rho$ decreases so the upper surface is setting $\rho = -1$, the middle sets $\rho=0$ and the lowest is for the case $\rho=+1$. When both revenues and costs are uncertain, their volatilities can still drive the value of delay, especially as $\rho$ decreases (which increases the volatility of profits).  When $\rho = 1$ the surface drops into a butterfly shape, i.e. the value of delay no longer increases with the volatilities \textit{per se} but with their difference. This is interesting for two reasons: first, by itself, more volatile revenues  need not increase the value of delay (and similarly, increasing cost uncertainty alone need not increase the value of delay); and second, this convex relationship between the value of delay and volatilities is due to revenue-cost correlation.  In the extreme case $\rho=1$  and $\sigma_1=\sigma_2$, the future project value becomes known with certainty (because all other parameters in \eqref{eqn.initial} are the same for revenues and costs) so the value of delay is zero. 

We compare Figure \ref{fig.volSf}{\color{red}(b)}, $r_p = 20\%, r_q = 30\%$ with  Figure \ref{fig.volSf}{\color{red}(c)}, $r_p = 30\%, r_q = 20\%$ In case (b) the investor requires a lower return than the market. If this is due to higher required return, that could be difficult to maintain when costs become highly uncertain, and this may be a reason why  cost volatility has a stronger impact on $v_0$ than the revenue volatility -- i.e. the surfaces are tilted.  If the higher return required by the investor is related to inefficiency in the use of capital, increasing uncertainty in costs has less effect.\footnote{The above results and more in the following, are generated with the same initial revenue and cost. In reality, they will diverge and naturally scale up or down these volatility effects. Which effect is more influential in determining $v_0$ is then an open question. For instance, when the expected revenue is high the NPV effect discussed in Observation \ref{sec.disagree} can dominate, causing an immediate investment in the project, whereas the cash flow volatilities lose their power to create any value of delay.}

\begin{figure}[!ht]\centering
\begin{minipage}{.45\linewidth}\centering
	\includegraphics[width=1\linewidth]{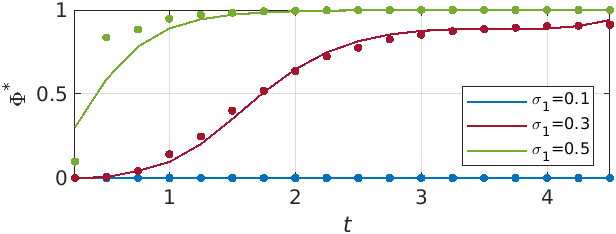}
\end{minipage}\quad
\begin{minipage}{.45\linewidth}{\centering
		\includegraphics[width=1\linewidth]{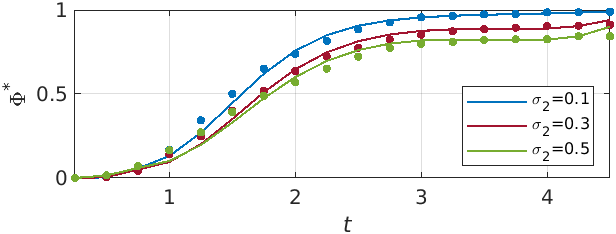}}
\end{minipage}
	\caption{\footnotesize 
	The probability of early exercise $\Phi^*$ for $\sigma_1$ (or $\sigma_2$) set at 10\%, 30\% and 50\%, with $r_p=0.28$ whilst $\sigma_2$ (or $\sigma_1$) and all other parameters as in \eqref{eqn.initial}. }\label{fig.EEBsigma}
\end{figure}

Figure \ref{fig.EEBsigma} shows that --  for an investor requiring a lower return than the market -- the probability of early exercise increases with the volatility of revenue and decreases with the volatility of cost, i.e. with the upside volatility in future project values. That upside dispersion drives the exercise decision has already been highlighted by \citet{CA2003,AS2004,BFJLR2008,CDRS2011,MCCC2011,F2015,MCP2018} and many others.

\section{Risk Analysis}\label{sec:hedging}
Classic real option valuation models \citep[\textit{e.g.}][and others]{G2011,AP2013,HHKN2016} rely on perfect hedging of all hedgeable risk factors involved in the investment. By contrast, as argued by \citet*{AF2006}, hedging practice is usually subjective and imperfect: it is really to avoid unwanted risks but not necessarily the mispriced ones (as perceived by the investor). In fact, if an investor disagrees with the market premium of a risk factor, they may not necessarily hedge it but attempt to exploit some abnormal return by retaining exposure to that risk factor, indeed by valuing the attached real option. Our SDCF appraoch fits into this paradigm very well, because we make no assumption about perfect, or even imperfect hedging. The investor is  free to hedge, or not, and if he does hedge then the value of the option to delay may decrease -- but not always. 

To see this, we now suppose $\mathcal{B}_0$ is a market-wide, hedgeable, systematic risk factor while the other two risk factors, $\mathcal{B}_1$ and $\mathcal{B}_2$, are idiosyncratic and un-hedgeable. This way, in the cash flow model \eqref{eqn.xcZB}, we decompose the investment risk into systematic and idiosyncratic components and examine how the option value depends on the factor premia required by the investor and the market. We assume that:
\begin{enumerate}\setlength{\itemsep}{-3pt}
	\item[(i)] 
Both cash flows are governed by identical parameters except their independent idiosyncratic factors. 
	That is, we set $\mu^*:=\mu_1=\mu_2$, $\sigma^*:=\sigma_1=\sigma_2$ and $\rho^* :=\rho_1=\rho_2$ in \eqref{eqn.xcZB}. This second assumption implies that the cost-revenue correlation is $\rho = {\rho^*}^2$;
	\item[(ii)]
The market expects a constant premium $\lambda_{0q}$ for the systematic risk factor $\mathcal{B}_0$, the investor expects a constant premium $\lambda_{0p}$ and we allow $\lambda_{0q} \ne \lambda_{0p}$. We assume the risk-free rate is zero, since it makes little difference to our sensitivity analysis. This way, the only determinants of the discount rates are the risk premia on different risk factors;
	\item[(iii)]
	Risk premia are also constant for  the two idiosyncratic risk factors $\mathcal{B}_1$ and  $\mathcal{B}_2$. Again, they can be  different for the investor and the market, but each has the same risk premium on both risk factors. We denote these premia as $\lambda_p$ for the investor and $\lambda_q$ for the market;
\end{enumerate}
Now suppose a proportion $h\in[0,1]$ of the systematic risk factor is hedged:
$h=0$ means no hedging at all and $h=1$ means perfect hedging. When there is hedging both the  cost and revenue volatilities change from $\sigma^*$ to another common value, viz.:
\begin{subequations}\label{eqn.hedge}
	\begin{equation}\label{eqn.hedgeMuSigma}
 \sigma_{h}=\sigma^*\sqrt{1-{(h \rho^*)}^2}\, .
	\end{equation}
	The discount rates of the investor and the market also vary with the hedging policy, and we can decompose them as a sum of two terms, the first representing the premia for the systematic (hedgeable) risk factor and the second being  premia for the idiosyncratic (un-hedgeable)  risk factors. That is:\footnote{Here the  risk-free rate is zero. Otherwise it would appear as a third term in \ref{eqn.hedgeR1}. We exclude it because it makes no qualitative difference to our conclusions.}
	\begin{equation}\label{eqn.hedgeR1}
	r_{ph}=\Big[\lambda_{0p}\big(1-h\big)\,\rho^*+\lambda_p\sqrt{1-{\rho^*}^2}\Big]\sigma^*\,,\quad r_{qh}=\Big[\lambda_{0q}\,\big(1-h\big)\,\rho^*+\lambda_q\sqrt{1-{\rho^*}^2}\Big]\sigma^*\,.
	\end{equation}
\end{subequations}

\subsection{{\normalfont\bf The hedging effect}\quad If the investor agrees with the  market on  the  systematic  risk premium then hedging has no effect on the value of delay. However, if they disagree then hedging reduces the value of delay. }\label{sec.hedge}

Observation \ref{sec.hedge} extends Observation \ref{sec.disagree} in that the disagreement effect also applies to any hedgeable risk factor involved in the investment. But only systematic risk can be reduced by hedging. This can reduce the investment option value, as well as the value of delay, whenever the investor disagrees with the market  on the factor premium; otherwise the option value is zero anyway and hedging will not change that.\footnote{This is just like hedging against a stock: if the investor believes that it is mispriced, then hedging simply eliminates the potential profit from exploiting the mispricing. Otherwise, any attempt to exploit any return more than expected by the market is pointless, let alone the timing of such an attempt.}

\begin{figure}[!ht]
	\centering
	\begin{minipage}{.49\linewidth}
		\includegraphics[width=1\textwidth]{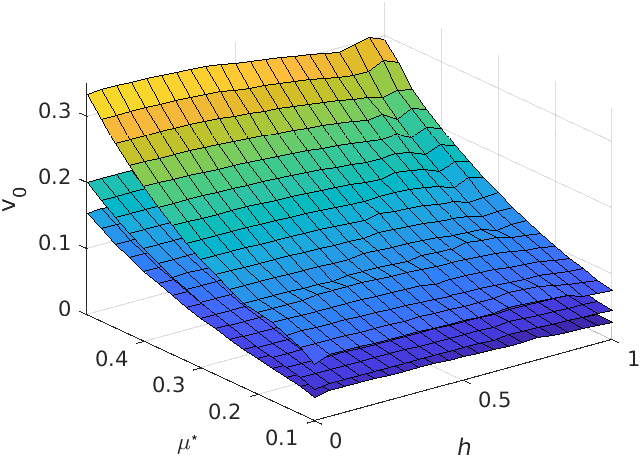}
	\end{minipage}
	\begin{minipage}{.49\linewidth}
		\includegraphics[width=1\textwidth]{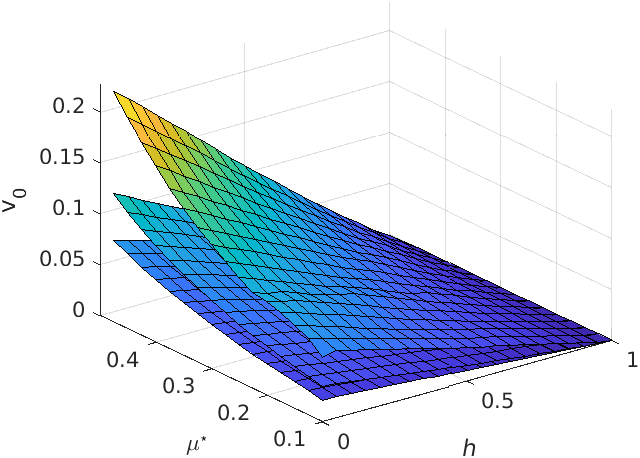}
	\end{minipage}
	\caption{\footnotesize The ROV with different values of $h$ and $\mu^*$.  The expected growth rates and volatilities of the  cash flows are calculated using (\ref{eqn.hedgeMuSigma}), and the implied and required returns are defined by (\ref{eqn.hedgeR1}). On the left is case (a), with $\lambda_{0p}=\lambda_{0q}=0.3$, and surfaces are drawn for $(\rho,\sigma) = (0.3,0.5), (0.3, 0.3), (0.6, 0.3)$ from highest to lowest. On the right is case (b) with $\lambda_{0q}=0.3$ and $\lambda_p=\lambda_q$ and surfaces are drawn for $(\rho,\sigma) = (0.3,0.5), (0.6, 0.3), (0.3, 0.3)$ from highest to lowest. In both, $r_p=28\%$ and $r_q=30\%$.}\label{fig.hedge}
\end{figure}

To quantify this effect, consider Figure \ref{fig.hedge}. In the left panel, case (a),  the  investor agrees with the market about the risk premium for $\mathcal{B}_0$ but disagrees on the premia for the other two factors, that is, $\lambda_{0p}=\lambda_{0q}$, $\lambda_{p}\neq\lambda_{q}$\,. The figure depicts the delay option value surfaces for a range of different values of $h$ and it clearly shows that the value of delay is independent of $h$, although different cash flow correlation and volatilities are used to compute the surfaces. In other words, the option value is unaffected by hedging the systematic risks associated with operating cash flows.  By contrast, the right panel (b) depicts the case that the investor disagrees with the market about the risk premium for $\mathcal{B}_0$ but agrees about the premia for the other two factors, \textit{i.e.} $\lambda_{0p}\neq\lambda_{0q}$, $\lambda_{p}=\lambda_{q}$. 
The right panel in Figure \ref{fig.hedge} depicts three option value surfaces  and all decrease with $h$, indicating a monotonic and negative impact of hedging on the option value. When $h=1$, the option value falls to zero regardless of other parameter values. 

Systematic risk factors may be  frequently traded, in which case  investors are less likely to consider them mispriced by the market. On the other hand, entrepreneurs seeking new business opportunities can exploit the market's mispricing of un-hedgeable idiosyncratic, and sometimes entirely new risks. Now we assume that the investor agrees with the market  on the systematic risk premium, so the option value stems only from un-hedgeable, idiosyncratic risks and their risk premia. Further, and following some classic asset pricing models  such as the CAPM model independently introduced by \citet{T1961,S1964,L1965,M1966}, and \citet{FF1993} three-factor model, we assume the market provides no risk premium for idiosyncratic risk factors. 
\subsection{{\normalfont\bf The idiosyncratic risk effect}\quad The value of delay is a concave function of idiosyncratic risk, especially when the investor requires a high return for bearing such a risk.}\label{sec.idio}
\begin{figure}[!ht]
	\centering
	\begin{minipage}{.45\linewidth}
		\includegraphics[width=1\linewidth]{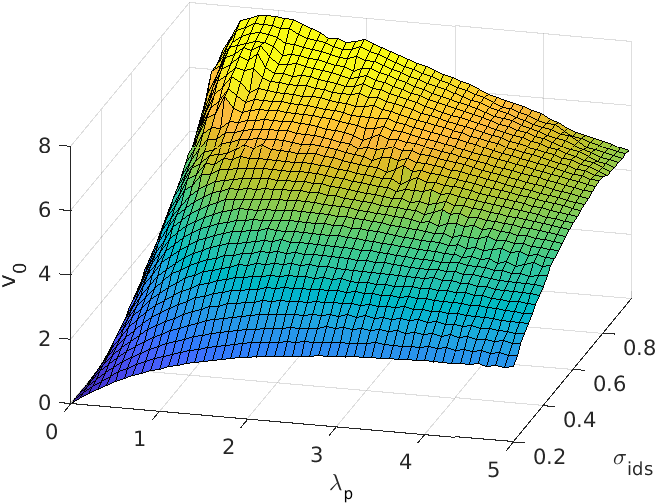}

	\end{minipage}
	\begin{minipage}{.45\linewidth}
	\includegraphics[width=1\linewidth]{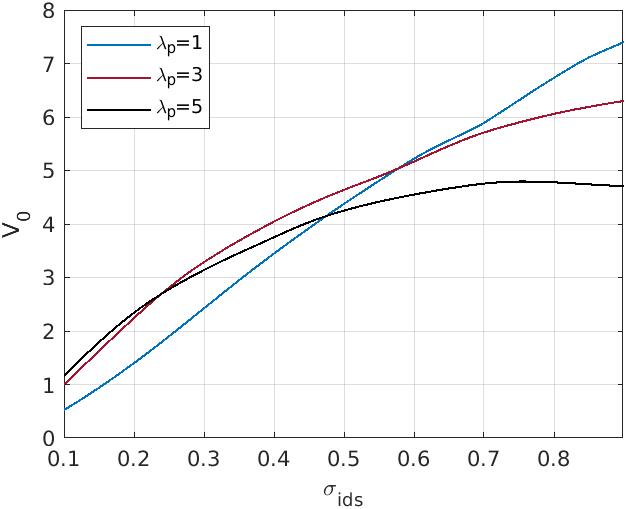}
	\end{minipage}
	\caption{\footnotesize  The value of delay for a range of idiosyncratic volatility $\sigma_{ids}$ and investor's idiosyncratic risk premia $\lambda_p$ , with $\lambda_{0p}=\lambda_{0q}=0.3$ and $\lambda_q=0$ in \eqref{eqn.hedgeR1}.  The figure on the right takes sections through the surface on the left at three different values for $\lambda_p$: the red, blue and black lines correspond to $\lambda_p=1,\,3,\,5$, respectively.}\label{fig.idiovol}
\end{figure}

To demonstrate this consider Figure \ref{fig.idiovol} which depicts the value of delay for different values of the idiosyncratic volatility $\sigma_{ids}=\sigma\sqrt{1-{\rho^*}^2}$ and of the risk premium perceived by the investor $\lambda_p$. 
The option value is positive everywhere because the investor can still perceive a positive future profit from investing later even though the market disagrees. Although $v_0$  increases with $\sigma_{ids}$ at low levels it starts to decrease when idiosyncratic risks are large. Especially when the investor requires  high premium on a high-risk factor, \textit{e.g.} when $\lambda_p=5$ and $\sigma_{ids}>0.7$ in the right panel, our SDCF approach confirms the findings of \citet{CKS2017} on the non-monotonic relationship between volatility and option values at high levels of risk. The economic rationale for this is that 
a highly risk-averse investor may require too much return for the profits available from un-hedgeable idiosyncratic risks, and therefore see limited value in the project not only now, but also in future. 

Assuming as we have that the investor agrees with the market on the systematic risk premium, and that the market risk premium for the idiosyncratic risk factor is zero, allows comparison of our results with those of  \citet{H2007} who focuses on the impact of idiosyncratic risk on the decision of a risk-adverse investor. When the investor and the market agree on a zero premium for bearing  idiosyncratic risk, both models produce an option value independent of other parameters. But in our case both the option value and the value of delay are zero. \citet{H2007} only considers the investors risk preferences but we also allow the cost of capital to influence discount rates and we model their divergence relative to the market.


\section{Conclusions}\label{sec.concl}
Modelling investor heterogeneity is very important, especially during periods of financial crises when there is likely to be much disagreement about expectations of market prices.  The required return on an investment is influenced by the investor's cost of capital, risk preferences, subjective beliefs on profitability of the investment and many other factors. By assuming no disagreement on the cash flow amounts, we emphasize how risk preferences in particular, but also the costs of capital can influence a subjective evaluation of an investment decision. 

By discounting the same cash flows at two different rates, respectively reflecting the investor and the market required returns, our SDCF model captures the subjective ROV arising from the investor's ability to exploit their perceived misvaluation by the market. We show that the value of delaying an investment increases as the investor's discount rate diverges from the market rate, and that the rate of this divergence drives the early exercise of the option. In other words, the greater the perceived mispricing of the project in its market value, the more subjective value the investor sees in the associated investment opportunity, and the faster the investor is to make the investment.  Another major difference between our SDCF model and other subjective models is that we address the limiting case of a zero subjective ROV when the investor and market agree on their required returns, in which case the investment decision opportunity is of no additional value to the investor. 

Importantly, by making no market completeness assumption we can model the impact of hedging on the subjective ROV, rather than requiring a specific hedging strategy for deriving the ROV as in both classic and utility-based approaches.
We show that the the effect of hedging  on the real option value is determined by the  investor's perceptions about mispricing in the market. If the investor and the market agree on a specific factor's risk premium then hedging this factor has no impact on the value of delay. Otherwise, it is the disagreement between systematic risk factors which drives the value of investment. Classic real option valuation models  rely on perfect hedging of all hedgeable risk factors and so they are not able to capture such effects.

Second, our model allows further decomposition of required returns into different risk premia, and this facilitates the analysis of impacts of various sources of risk on the subjective ROV. Idiosyncratic risks are heightened during crisis periods, when it is particularly important to understand their effect on investment valuations. 
Assuming no disagreement between the investor and the market on  systematic (i.e. hedgeable) risks, we demonstrate that the perceived value of the option to invest, now or in the future, tends to increase with the investor's idiosyncratic (i.e. un-hedgeable) risks associated with it. Unless the investor is highly risk averse -- or highly inefficient in his use of capital -- he could perceive a positive value in bearing idiosyncratic risks even though the market attributes a zero risk premium to such risks. Delaying investment becomes more valuable but decisions maker also tend to rush their decisions when idiosyncratic risks become exceptionally high. This, to some extent, explains the observations made by \citet{liu2021investment} on the market aggregate investment level decreasing with idiosyncratic risks and increasing when the risk level continues to rise.

A numerical example compares our results with those of the \cite{CA2003} based on identical inputs. { The SDCF investment real option values are smaller than the MAD option values, and likewise for the value of the option to delay investment. An Excel workbook allows readers to define their own scenarios for comparison. 	Then, through simple discounting of cash flows, we capture some novel features that have previously been recorded only in complex models that are difficult to implement in practice. These simulations assume two correlated GBM for the operating revenues and costs, and to generate the stylised facts described above we also assume, for simplicity, that the discount rates for operating revenues and costs  are  equal. However, as the reader can verify using the Python code provided, the observations made in this article are quite general and are not restricted by these assumptions.}

Many of our results are new, although some are simply verifying features that are already well-established in the real option literature. For instance, 
we quantify how the subjective real option value increases with the volatility of project values, a general feature which is already well known from classic models.
Overall, our most significant contribution is to present a model that is very easy to apply in practice, replicates features already established by more restrictive models, establishes important new results, and requires no more data than traditional DCF approaches.

\newpage
\addcontentsline{toc}{section}{References}
\fancyhead[R]{}
\spacing{1}\bibliography{Risk-Adjusted_Valuation_for_Real_Option_Decisions}
\bibliographystyle{apalike}
\newpage
\begin{appendices}
\section{Appendix: Deriving $\mbox{NPV}_t^0$ }\label{app.npv}
We derive \eqref{eqn.po} from \eqref{eqn.subj} to \eqref{eqn.NPVQ} with correlated revenues and costs. Following our assumption of the revenue and cost evolutions \eqref{eqn.xcZB}, we can write:
$$\mathbb{E}_t[x_\tau] = x_{1t}\exp(\mu_1(\tau-t))-x_{2t}\exp(\mu_2(\tau-t))$$
and therefore, 
\begin{align}\nonumber
p_t& =\int^T_t\left(x_{1t}\exp(\mu_1(\tau-t))-x_{2t}\exp(\mu_2(\tau-t))\right)\exp(-r_p(\tau-t))d\tau\\\nonumber &=x_{1t}\int^T_t\exp((\mu_1-r_p)(\tau-t))d\tau-x_{2t}\int^T_t\exp((\mu_2-r_p)(\tau-t))d\tau\\\nonumber
&= x_{1t}\frac{\exp((\mu_1-r_p)(T-t))-1}{\mu_1-r_p} - x_{2t}\frac{\exp((\mu_2-r_p)(T-t))-1}{\mu_1-r_p}\,,
\end{align}
and similarly,
$$q_t = x_{1t}\frac{\exp((\mu_1-r_q)(T-t))-1}{\mu_1-r_q} - x_{2t}\frac{\exp((\mu_2-r_q)(T-t))-1}{\mu_2-r_q}\,.$$
Thus,
\begin{align}\nonumber
\mbox{NPV}_t=&\, p_t-q_t\\\nonumber
 =&\, x_{1t}\left[\frac{\exp((\mu_1-r_p)(T-t))-1}{\mu_1-r_p}-\frac{\exp((\mu_1-r_q)(T-t))-1}{\mu_1-r_q}\right] \\\nonumber
&-x_{2t}\left[\frac{\exp((\mu_2-r_p)(T-t))-1}{\mu_2-r_p}-\frac{\exp((\mu_2-r_q)(T-t))-1}{\mu_2-r_q}\right],
\end{align}
and so,
\begin{align}\nonumber
\mbox{NPV}_t^0=&\, (p_t-q_t)\exp(-r_pt)\\\nonumber
=&\, x_{1t}\left[\frac{\exp((\mu_1-r_p)(T-t))-1}{\mu_1-r_p}-\frac{\exp((\mu_1-r_q)(T-t))-1}{\mu_1-r_q}\right]\exp(-r_pt) \\\nonumber
&-x_{2t}\left[\frac{\exp((\mu_2-r_p)(T-t))-1}{\mu_2-r_p}-\frac{\exp((\mu_2-r_q)(T-t))-1}{\mu_2-r_q}\right]\exp(-r_pt).\\\nonumber
=&\,c_{1t}\,x_{1t}-c_{2t}\,x_{2t}\,,
\end{align}
where $\displaystyle c_{it} = \left[\frac{\exp((\mu_i-r_p)(T-t))-1}{\mu_i-r_p}-\frac{\exp((\mu_i-r_q)(T-t))-1}{\mu_i-r_q}\right]\exp(-r_pt)$.

\end{appendices}
\end{document}